\documentclass[a4paper,twocolumn,11pt,accepted=2024-12-12]{quantumarticle}
\pdfoutput=1
\usepackage[utf8]{inputenc}
\usepackage[english]{babel}
\usepackage[T1]{fontenc}
\usepackage{amsmath,bm}
\usepackage{amsfonts}
\usepackage{graphicx}
\usepackage{subfigure}
\usepackage{braket}
\usepackage{hyperref}
\usepackage{tikz}
\usepackage{lipsum}
\usepackage{subfiles}
\usepackage[numbers,sort&compress]{natbib}
\usepackage{comment}
\usepackage{xtab}

\begin{document}

\title{Low-Rank Variational Quantum Algorithm for the Dynamics of Open Quantum Systems}
\author[1,2]{Sara Santos}
\email{sara.alvesdossantos@epfl.ch}
\author[1,2]{Xinyu Song}
\email{xinyu.song@epfl.ch}
\author[1,2]{Vincenzo Savona}
\email{vincenzo.savona@epfl.ch}

\affil[1]{Institute of Physics, Ecole Polytechnique Fédérale de Lausanne (EPFL), CH-1015 Lausanne, Switzerland}
\affil[2]{Center for Quantum Science and Engineering, Ecole Polytechnique Fédérale de Lausanne (EPFL), CH-1015 Lausanne, Switzerland}
\maketitle
\begin{abstract}
    \noindent The simulation of many-body open quantum systems is key to solving numerous outstanding problems in physics, chemistry, material science, and in the development of quantum technologies. Near-term quantum computers may bring considerable advantage for the efficient simulation of their static and dynamical properties, thanks to hybrid quantum-classical variational algorithms to approximate the dynamics of the density matrix describing the quantum state in terms of an ensemble average. Here, a variational quantum algorithm is developed to simulate the real-time evolution of the density matrix governed by the Lindblad master equation, under the assumption that the quantum state has a bounded entropy along the dynamics, entailing a low-rank representation of its density matrix. The algorithm encodes each pure state of the statistical mixture as a parametrized quantum circuit, and the associated probabilities as additional variational parameters stored classically, thereby requiring a significantly lower number of qubits than algorithms where the full density matrix is encoded in the quantum memory. Two variational ansatze are proposed, and their effectiveness is assessed in the simulation of the dynamics of a 2D dissipative transverse field Ising model. The results underscore the algorithm's efficiency in simulating the dynamics of open quantum systems in the low-rank regime with limited quantum resources on a near-term quantum device.
\end{abstract}

\section{Introduction}

Quantum systems are invariably influenced by their environment~\cite{Breuer_2007}, an interaction that often has detrimental effects such as decoherence, thereby limiting the efficacy of quantum technological platforms. The ability to efficiently simulate open quantum systems is thus of paramount importance, providing essential insights into their fundamental properties and aiding in the development of quantum technologies. 

The statistical properties of an open quantum system are fully described by the density matrix, which gives direct access to statistical ensemble averages of physical quantities. The time evolution of the density matrix is governed by a master equation which, within the general assumption of a Markovian environment, takes the celebrated Lindblad form~\cite{Breuer_2007}. Alternatively, open quantum systems admit a stochastic description in terms of \textit{quantum trajectories}, which simulate their dynamics by unraveling the evolution of the system's density matrix into an ensemble of stochastically evolving pure states~\cite{Daley_2014,Dum_1992,Dalibard_1992,charmichael_1991}.
A variety of numerical methods have been developed for the simulation of open quantum system dynamics, either through direct integration of the master equation or by averaging over simulated quantum trajectories~\cite{WeimerRMP2021}. One of the main challenges in the simulation of open quantum systems is the increased computational complexity due to the mixed nature of the quantum state. If the unitary dynamics of an $n$-qubit system is described by the time-evolution of its $N=2^n$ amplitudes, a mixed state of the same system requires expressing $N^2$ elements of the density matrix. Simulating one quantum trajectory has a computational complexity comparable to pure quantum states, but averaging over many trajectories is required to estimate physical quantities with sufficient accuracy. Finally, it has been shown~\cite{PreisserPRA2023} that the entanglement entropy of the density matrix grows in time significantly more slowly than that of individual quantum trajectories, thereby making the approximate numerical description more efficient. 

Quantum computers excel at representing quantum states efficiently, offering the potential to simulate the real-time evolution of quantum systems, as initially proposed by Richard Feynman~\cite{feynman}. Seth Lloyd has shown that the time evolution operator $\hat{G}(t)=e^{-i\hat{H}t}$ can be implemented on quantum computers efficiently when the Hamiltonian $\hat{H}$ of the system only carries local interaction~\cite{doi:10.1126/science.273.5278.1073}. In particular, for $k$-local Hamiltonians~\cite{tacchino_2020}, the operator $\hat{G}$ can be efficiently approximated via the Trotter-Suzuki decomposition~\cite{tacchino_2020,Childs_RMP_2021}, leading to a polynomial gate complexity in the system size~\cite{tacchino_2020}. However, near-term, a.k.a NISQ devices~\cite{Preskill2018quantumcomputingin}, with limited coherence time and operation accuracy, put constraints on the application of Suzuki-Trotter time-evolution schemes, since the required quantum circuit depth scales proportionally to the number of time steps $N_t$.

To address the limitations of NISQ devices, variational quantum algorithms (VQAs), have emerged~\cite{cerezo_2021,Bauer_Algorithms_2020,Bharti_Noisy_RMP_2022}. VQAs efficiently prepare quantum states as low-depth parameterized quantum circuits and evolve/optimize the parameters to capture the non-equilibrium dynamics of a given physical system. In the context of time evolution, variational methods have the advantage of a significantly reduced circuit depth and gate count, thanks to the availability of circuit-efficient variational ansatze~\cite{cerezo_2021}, making it possible to approximately simulate the unitary evolution with shallow circuits on NISQ devices~\cite{Barison2021efficientquantum,Benjamin_2017}.

Lindblad dynamics, however, is associated to a non-unitary time-evolution operator, which cannot be directly encoded as a quantum circuit. Several quantum algorithms for simulating the dynamics of open quantum systems have been proposed~\cite{Barreiro_2011,wang_2011,sweke_2014,sweke_2015,Di_Candia_2015,sweke_2016,chenu_2017,cleve2019efficient,Garcia_Perez_2020,su_2020,Endo_2020,PhysRevResearch.2.043289,Ramusat2021quantumalgorithm,PhysRevLett.127.020504,PRXQuantum.3.010320,Liu2022,Kais_2022,watad2023variational,Chen2024adaptivevariational}. These are mostly based on an appropriate dilation of the system's Hilbert space to encode environment degrees of freedom, and on mid-circuit measurements and/or state re-initializations to model dissipative processes, or rely on the Suzuki-Trotter expansion of the non-unitary dynamics~\cite{PhysRevLett.127.020504,PRXQuantum.3.010320}. 

Recently, variational quantum algorithms have been introduced to simulate the Lindblad master equation directly, relying on a full encoding of the density matrix as a parametrized circuit. For a system whose pure quantum states can be encoded on $n$ qubits, these algorithms typically require encoding the density matrix on a $2n$-qubit register~\cite{PhysRevResearch.2.043289,watad2023variational,Chen2024adaptivevariational}, while deep quantum neural-networks ansatze can have a lower, but still $>n$, qubit requirement~\cite{Liu2022}. Similar parameterized ansatze for learning mixed and thermal states~\cite{terhal2000,Ezzell_2023,PhysRevA.101.012328,PhysRevB.106.165126} have also been studied.

In a variational approach to time evolution, a quantum circuit is executed only to estimate a single timestep, while the dependence on the number $L$ of operators associated to the dissipation processes, and the system size $n$, are embedded in the representative power of the variational ansatz being used. To achieve constant accuracy on the other hand, non-variational algorithms require quantum circuits whose depth grows linearly as $N_t\times L\times n$~\cite{tacchino_2020}, with the additional caveat that ancilla qubits and noisy non-local multi-qubit gates are usually needed. This is the actual source of the potentially very significant gate complexity advantage of variational algorithms, leading to the possibility to run them on NISQ devices, while the price to be paid is in the large number of circuits executions required by a VQA to estimate expectation values.

Under several common conditions, open quantum systems are characterized by low entropy, corresponding to quantum states with a low degree of mixedness. Typical cases are the transient dynamics starting from an initial pure state, weak system-bath coupling, as well as many gate and readout protocols on quantum information platforms, which are designed to sustain nonclassical states as long as possible, thereby featuring slow mixing and entropy growth rates. A paradigmatic example is the control of qubits, such as the transmon or the Schrödinger cat qubit, which are characterized by low entropy~\cite{blais_2023_transmon_chaos,gautier2024optimalcontrollargeopen,gravina2023adaptive} along the dynamics of the main readout and gate protocols. More generally, it is expected that the entropy of an open quantum system becomes extensive (in the system size) only in the vicinity of dissipative phase transitions~\cite{Donatella_2021, Finazzi_2015_corner, LeBris_2013_low_rank, McCaul_2021}, where the Liouvillian gap closes~\cite{znidaric_2015,Kastoryano_2013}, while it stays limited in the vast majority of other cases.

A low-entropy quantum state admits a low-rank approximation, by retaining only the dominant density matrix eigenvalues. The low-rank approximation is therefore a valuable opportunity to significantly reduce the computational complexity of the simulation. Several low-rank approaches have been recently developed for the simulation on classical computers~\cite{Donatella_2021,Ezzell_2023,https://doi.org/10.1038/s41534-020-00302-0}, including a time-dependent variational algorithm with adaptive dynamical rank~\cite{gravina2023adaptive}. Given the broad range of validity of the low-rank approximation, a quantum algorithm for the dynamics of low-rank open quantum systems would bring a considerable advantage to the efficient simulation of these systems.

Here, we develop a variational quantum algorithm that simulates the dynamics of open quantum systems with a truncated rank parameterized density matrix. The mixed state is represented as a statistical mixture of pure states, and each pure state is encoded as a parametrized quantum circuit, while the probabilities, also entering as variational parameters, are stored classically, taking advantage of the low-rank assumption of $\mathcal{O}(\mathrm{poly(n)})$ nonzero probabilities. The main advantage of this representation is that the required quantum resources are the same as for a pure quantum state, e.g. $n$ qubits for a system of $n$ interacting spins, as opposed to the $2n$ qubits that would be required to encode the most general density matrix. We introduce two variational ansatze, characterized by different trade-offs between expressive power and required computational resources.

We demonstrate the algorithm on a 2D dissipative transverse field Ising model (TFIM) and discuss the practical aspects of implementing it on a quantum device. Specifically, we compute the computational cost of implementing the two ansatze in terms of rank, the system size, and we assess the depth, gate count, and number of variational parameters necessary to achieve a desired accuracy. Furthermore, we evaluate the algorithms performance in noisy quantum hardware when simulating the same physical model. 

The article is structured as follows. In Section \ref{methods}, we review the theoretical background for Markovian open quantum systems and variational quantum time evolution, and we introduce the quantum algorithm with the two low-rank parametrized representations of the quantum state. In Section \ref{results}, we benchmark and compare the performance of two different ansatze, both within statevector simulation and with Qiskit's Qasm simulator including noise. Section \ref{conclusion} presents our conclusions and the outlook of our work.
\section{Model and methods}
\label{methods}
The dynamics of an open quantum system interacting with a Markovian environment is governed by the Lindblad master equation~\cite{Breuer_2007}
\begin{align}
\dot{\hat{\rho}} &= \mathcal{L}\left[\hat{\rho}\right]=-i\left[\hat{H},\hat{\rho}\right]\nonumber\\
&+\sum_{k}\gamma_k\left(\hat{c}_k\hat{\rho} \hat{c}_k^{\dag}-\frac{1}{2}\left\{\hat{c}_k^{\dag}\hat{c}_k, \hat{\rho}\right\}\right)\,, \label{eq:1}
\end{align}
where $\mathcal{L}$ is the Liouvillian superoperator, $\hat{\rho}$ is the density matrix of the system, $\hat{H}$ is the system's Hamiltonian, $\hat{c}_k$ is the $k-th$ jump operator which describes the mechanism of dissipation, and $\gamma_k$ is the corresponding dissipation rate. For most systems of physical interest, both $\hat{H}$ and $\hat{c}_k$ are quasi-local operators, i.e., they can be expressed as sums of low-weight Pauli strings, whose number scales polynomially with the system size $n$ ~\cite{kempe2005complexity}. 

\subsection{Low-rank diagonal ansatz}
\label{ansatz}
\begin{figure*}[t!]
    \centering
    \includegraphics[width=\textwidth]{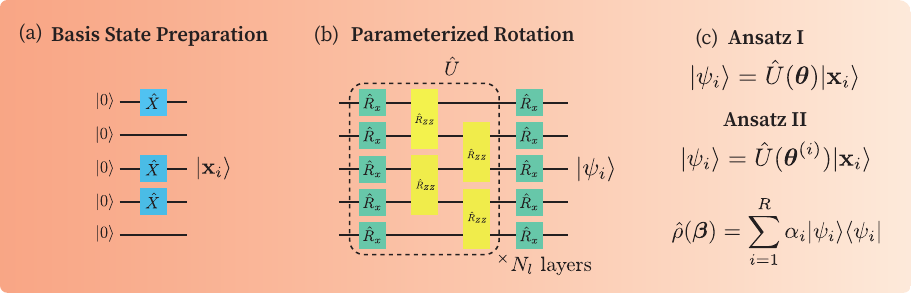}
    \caption{\textbf{Encoding the variational ansatz in a quantum processor.} (a) The computational basis states $|\mathbf{x}_i\rangle$ are prepared in a quantum circuit by application of Pauli $\hat{X}$ gates. (b) Each state $|\mathbf{x}_i\rangle$ is then rotated according to a sequence of parameterized gates $\hat{U}$, consisting of $N_{l}$ layers of single qubit rotations and multi-qubit entangling gates, generating the pure states $|\psi_i\rangle$ which enter the statistical mixture. (c) The variational ansatz $\hat{\rho}(\boldsymbol{\beta})$ is expressed as a linear combination of pure states, where each computational basis state $|\mathbf{x}_i\rangle$ is rotated by the same unitary $\hat{U}(\boldsymbol{\theta})$ - ansatz I, or individually by $\hat{U}(\boldsymbol{\theta}^{(i)})$ - ansatz II.}
    \label{fig:scheme_ansatz1}
\end{figure*}

A density matrix can always be expressed in diagonal form as
\begin{equation}
    \hat{\rho}=\sum_{i=1}^N p_i\ket{\psi_i}\bra{\psi_i}\,,
\end{equation}
where $\ket{\psi_i}$ are a set of not necessarily orthogonal but linearly independent states. In this work, we assume that the open quantum system is characterized by low entropy along its dynamics, it is convenient to approximate it in terms of a truncated-rank ansatz as~\cite{Ezzell_2023,verdon2019quantum,Liu2021}
\begin{equation} \label{eq:a1}
\hat{\rho}(\bm{\beta}) = \sum_{i=1}^R \alpha_{i} \hat{U}(\bm{\theta})\ket{\mathbf{x}_i} \bra{\mathbf{x}_i}\hat{U}(\bm{\theta})^{\dag}\,,
\end{equation}
where $\mathbf{x}_i$ are a set of states selected among the $2^n$ computational basis states of $n$ qubits, 
$\bm{\beta} = (\bm{\alpha},\bm{\theta})$ are variational parameters, $\alpha_i\in\mathbb{R}^+$, $\hat{U}(\bm{\theta})$ is a parameterized unitary operator which is encoded as a variational quantum circuit. This ansatz does not assume $\mathrm{Tr}(\hat{\rho})=1$, which must be taken into account when estimating expectation values of operators. For this variational ansatz to be efficient, we need the rank $R$ to be $\mathcal{O}(\mbox{poly}(n))$. This condition is often fulfilled in many systems of physical interest~\cite{McCaul_2021,Donatella_2021,chen2020low,LeBris_2013_low_rank,gravina2023adaptive}.
A possible, and arguably more expressive generalization of ansatz \eqref{eq:a1}, consists in introducing a different set of values for the parameters $\bm{\theta}$ for each state in the mixture, resulting in
\begin{equation} \label{eq:a2}
\hat{\rho}(\bm{\beta}) = \sum_{i=1}^{R} \alpha_i \hat{U}(\bm{\theta^{(i)}})\ket{\mathbf{x}_i} \bra{\mathbf{x}_i}\hat{U}(\bm{\theta^{(i)}})^{\dag}\,.
\end{equation}
In what follows, we refer to~\eqref{eq:a1} and \eqref{eq:a2} as ansatze I and II respectively. Notice that, while in ansatz I the states of the statistical mixture are mutually orthogonal, this is not true in ansatz II. The two ansatze are summarized in the diagram presented in Fig.~\ref{fig:scheme_ansatz1}. 
Here we adopt, for the parameterized circuit $\hat{U}$, a hardware efficient parametrized circuit~\cite{cerezo_2021,Bharti_Noisy_RMP_2022,Bauer_Algorithms_2020}, with $N_{l}$ layers of single- and multiple-qubit Pauli rotation gates. The gate complexity of $\hat{U}$ scales linearly with the number of independent non-commuting terms of the Hamiltonian, which is typically of the same order of magnitude as the system size. The expressive power of ansatz I and ansatz II depends on both $N_{l}$ and $R$, which measure respectively the amount of entanglement and the mixedness of the quantum state. It has been suggested that in physically relevant systems there is a trade-off between these two features, as intuitively the influence of the environment partially suppresses the entanglement of the system. Here, however, we will study the effectiveness of each ansatz as a function of the two parameters independently. 
Ansatze I and II have different computational costs, with ansatz II requiring a larger number of parameters and measurements in the estimate of expectation values. In Appendix \ref{Cost}, we analyze the resource requirements of the present algorithm and show that the number of quantum circuits to be measured scales as $\mathcal{O}(LR^2N_{\theta}+RN_{\theta}^2)$ and $\mathcal{O}(LR^2 + LRN_{\theta} + N_{\theta}^2)$ for ansatze I and II, respectively, with $N_{\theta}$ the total number of angular parameters.
\subsection{McLachlan's variational principle}
We review the essential aspects of McLachlan's variational principle~\cite{Yuan2019theoryofvariational}. Assume a general variational quantum state $\hat{\rho}(\bm{\beta}(t))$, where $\bm{\beta}$ is a set of real-valued parameters. Evolution over a small time interval $\delta t$ is described by a variation of $\bm{\beta}$ defined by
\begin{align}
\hat{\rho}(\bm{\beta}(t+\delta t)) &= \hat{\rho}(\bm{\beta}(t)) \nonumber \\
&+ \sum_j \frac{\partial \hat{\rho}(\bm{\beta})}{\partial \beta_j}\dot{\beta_j}\delta t + \mathcal{O}(\delta t^2)\,.  \label{rho_I}
\end{align}
The same evolution, according to the master equation Eq.~\eqref{eq:1}, is instead given by
\begin{align}
\hat{\rho}(\bm{\beta}(t+\delta t))&=\hat{\rho}(\bm{\beta}(t))\nonumber\\
&+ \mathcal{L}\left[\hat{\rho}(\bm{\beta}(t))\right] \delta t + \mathcal{O}(\delta t^2)\,.  \label{rho_II}
\end{align}
The McLachlan's variational principle follows from the minimization of the $L^2$-distance between states~\eqref{rho_I} and~\eqref{rho_II}. To leading order in $\delta t$, this corresponds to minimizing the norm $\left\lVert \sum_j \frac{\partial \hat{\rho}(\bm{\beta})}{\partial \beta_j}\dot{\beta_j} -  \mathcal{L}\left[\hat{\rho}(\bm{\beta})\right] \right\rVert$ with respect to $\dot{\bm{\beta}}$. This leads to the set of differential equations
\begin{equation} \label{eq:MV_MA}
\begin{aligned}
&\sum_j M_{kj}\dot{\beta_j} = V_k\, ,\\
&M_{kj} = \mathrm{Tr} \left[\frac{\partial \hat{\rho}(\bm{\beta})}{\partial \beta_k} \ \frac{\partial\hat{\rho}(\bm{\beta})}{\partial \beta_j}\right]\,,\\
&V_k =\mathrm{Tr} \left[\frac{\partial\hat{\rho}(\bm{\beta})}{\partial \beta_k} \ \mathcal{L} \left[\hat{\rho}(\bm{\beta})\right]\right],
\end{aligned}
\end{equation}
where $M_{kj}$ are the elements of a positive semi-definite matrix $M$, and $V_k$ the components of an array $V$. 
The variational quantum time evolution algorithm can be summarized as follows. For each time step, the entries of $M$ and $V$ are computed by estimating terms of the form

\begin{align*}
   & \bra{\psi_i}\hat{O}\ket{\psi_j}\,,~\bra{\partial\psi_i/\partial\theta_k}\hat{O}\ket{\psi_j}\,,\\
   & \bra{\psi_i}\hat{O}\ket{\partial\psi_j/\partial\theta_k}\,,~\bra{\partial\psi_i/\partial\theta_k}\hat{O}\ket{\partial\psi_j/\partial\theta_k}\,,
\end{align*}
on a quantum device, where $\hat{O}$ is a unitary operator typically given by a low-weight Pauli string, and $\ket{\psi_i}=\hat{U}(\bm{\beta})\ket{\mathbf{x}_i}$ and $\ket{\psi_i}=\hat{U}(\bm{\beta}^{(i)})\ket{\mathbf{x}_i}$, for ansatze I and II, respectively. Then, the linear system Eq.~\eqref{eq:MV_MA} is solved on a classical processor and the variational parameters are updated. This step is iterated until the final time is reached. The expressions of Eqs.~\eqref{eq:MV_MA} and of the quantum circuits are detailed in Sections \ref{EOMI} and \ref{EOMII} of the Appendix.
Note that the present formulation is not trace-preserving. Previous studies ~\cite{10.1063/1.4916384} have addressed this issue by enforcing trace conservation with Lagrange multipliers. Here, we adopt an approach consisting in estimating expectation values of observables $\hat{O}$ as
\begin{equation}
\langle \hat{O}\rangle = \frac{\mathrm{Tr}[\hat{\rho}(\bm{\beta})\hat{O}]}{\mathrm{Tr}[\hat{\rho}(\bm{\beta})]}= \frac{\sum_i \alpha_i \bra{\psi_i}\hat{O}\ket{\psi_i}}{\sum_i \alpha_i}\,.
\end{equation}
Finally, it is important to notice that the present algorithm requires only $n$ qubits, i.e., as many as are needed to encode the pure states $\ket{\psi_i}$. This comes as a considerable advantage compared to most algorithms relying on the purification of the quantum state~\cite{Yuan2019theoryofvariational} or vectorization of the density matrix~\cite{PhysRevResearch.2.043289}, which instead require $2n$ qubits. This advantage is particularly relevant in view of the application of error mitigation strategies~\cite{cai2023quantum,Mari_2021,berg2022probabilistic,Endo_2018,Endo_2021,Nation_2021,Yang_2022} whose effectiveness depends critically both on the width and depth of a quantum circuit.

\section{Numerical Results}
\label{results}
In this section, we demonstrate the low-rank quantum time evolution (LRQTE) algorithm on the dissipative TFIM. This model is characterized by the Hamiltonian
\begin{equation} \label{eq:16}
\hat{H}_{TFI} = J_z\sum_{\langle j,k\rangle}\hat{\sigma}_j^z \hat{\sigma}_k^z + h\sum_j \hat{\sigma}_j^x \, ,
\end{equation}
and by the dissipative dynamics generated by the jump operators $\hat{c}_k = \hat{\sigma}_k^{-}=\frac{\hat{\sigma}_k^x-i\hat{\sigma}_k^y}{2}$. Here, $\hat{\sigma}_j^\alpha$ are Pauli matrices and we assume a lattice of interacting spin-1/2 sites, where $J_z$ is the coupling strength, $h$ is an external magnetic field, and $\langle j,k\rangle$ indexes pairs of nearest neighbouring sites. We consider the case with $J_z=1$, $h=0.5$, and $\gamma_k = \gamma = 1$. We further assume the initial state to be $\ket{\downarrow \downarrow ... \downarrow}\bra{\downarrow \downarrow ... \downarrow}$, i.e., the all-spin down configuration. With this choice of system parameters and initial state, the Lindblad dynamics towards the steady state is low-rank, as will appear from the simulations. 
As a proof-of-concept, we present the numerical results of the noise-free statevector simulation with Julia's Yao package~\cite{YaoFramework2019} of a $3\times3$ square lattice with open boundary conditions, which we use as a benchmark for the algorithm. Specifically, we compute the average magnetization along the $z$-axis $s_z=\mathrm{Tr}[\hat{\rho}\hat{\sigma}_z]$, with $ \hat{\sigma}_z=\sum_{i=1}^n \hat{\sigma}_i^z/n$, the average magnetization along the $x$-axis $s_x=\mathrm{Tr}[\hat{\rho}\hat{\sigma}_x]$, with $\hat{\sigma}_x =\sum_{i=1}^n \hat{\sigma}_i^x/n$, the purity of the variational state $P=\mathrm{Tr}\left [\hat{\rho}^2\right]$, and the infidelity $1-\mathcal{F}$, where $\mathcal{F}=\mathrm{Tr} \left[ \sqrt{\sqrt{\hat{\rho}}\hat{\rho}_0\sqrt{\hat{\rho}}} \right]^2$ is the fidelity computed with respect to the exact time-evolved state $\hat{\rho}_0$ obtained with the package QuTiP~\cite{Johansson_2012,Johansson_2013} from the direct integration of the Lindblad master equation. We consider here a problem-inspired parametrized quantum circuit $\hat{U}(\bm{\theta})$~\cite{Barison2021efficientquantum}
\begin{equation}\label{eq:ising_trotter}
\begin{split}
    \hat{U}(\bm{\theta}) &= \prod_{l=1}^{N_{l}}\hat{U}_{l}(\theta_l)\\
    &=\prod_{l=1}^{N_{l}}\left[\prod_{j=1}^n e^{-i\hat{\sigma}_j^{x}\frac{\theta_{j,l}}{2}}\right]\left[\prod_{\langle j,k\rangle}e^{-i\hat{\sigma}_j^z\hat{\sigma}_k^z\frac{\theta_{\langle j,k\rangle,l}}{2}}\right] \, .
\end{split}
\end{equation}
We compare the accuracy and the resource efficiency of the two variational ansatze, and investigate the importance of the choice of the truncated basis for the LRQTE. In addition, a simulation of a smaller instance of the one-dimensional dissipative TFIM is carried out on Qiskit's Qasm Simulator~\cite{gadi_aleksandrowicz_2019_2562111}, accounting for noise and the connectivity of IBM's Lagos quantum device.
\begin{figure}[h!]
    \centering
    \includegraphics[width=\linewidth]{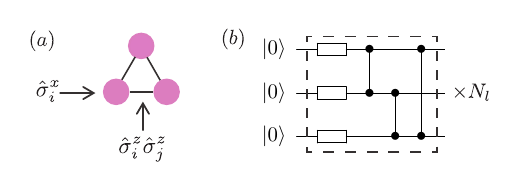}
    \caption{\textbf{Ansatz construction.} (a) 3 qubit TFIM. The circles represent the qubits and the lines the nearest-neighbor interactions. (b) Parameterized circuit $\hat{U}$, composed of $N_{l}$ repetitions of single-qubit $\hat{R}_x$ gates and two-qubit $\hat{R}_{zz}$ gates (see Eq.\eqref{eq:ising_trotter}). The total number of parameterized gates is $N_{l}(n+E)$, with $E$ the number of $\langle i,j\rangle$ pairs of nearest neighbouring sites.}
    \label{fig:ising_ansatz}
\end{figure}

\subsection{Statevector Simulation}\label{sec:SV}
  \begin{figure}[h!]
    \centering
    \includegraphics[width=0.5\textwidth]{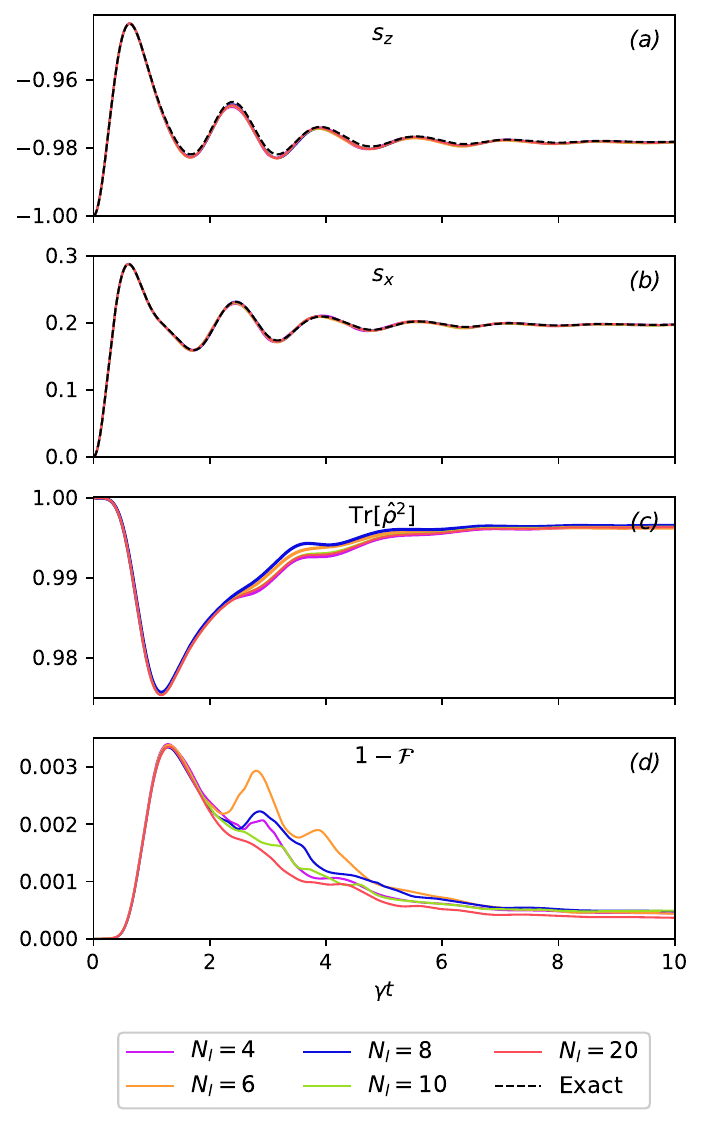}
    \caption{Time evolution of a $3\times3$ dissipative TFIM with $J_z=\gamma=1$, $h=0.5$ and $dt=0.005$, simulated with ansatz I with $\textrm{rank} =10$: (a) average magnetization along the $x$-axis $s_x$, (b) average magnetization along the $z$-axis $s_z$, (c) purity of the variational state $\hat{\rho}$, (d) infidelity $1-\mathcal{F}$ between the variational and exact quantum states.}
    \label{Fig.SV_ansatz1}
\end{figure}
In Figs.~\ref{Fig.SV_ansatz1} and \ref{Fig.SV_ansatz2} we present the results of the noise-free statevector simulation using ansatze I and II, respectively.
    \begin{figure}[h!]
    \centering
    \includegraphics[width= 0.5\textwidth]{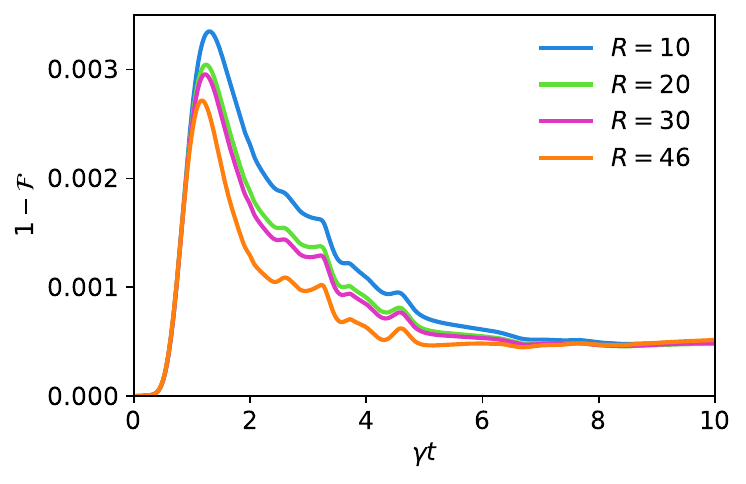}
    \caption{Infidelity between the variational and exact quantum states, computed with ansatz I with 10 layers and rank $R=\{10,20,30,46\}$.}
    \label{Fig.SV_ansatz1_rank}
\end{figure}
    \begin{figure}[h!]
    \centering
    \includegraphics[width=0.5\textwidth]{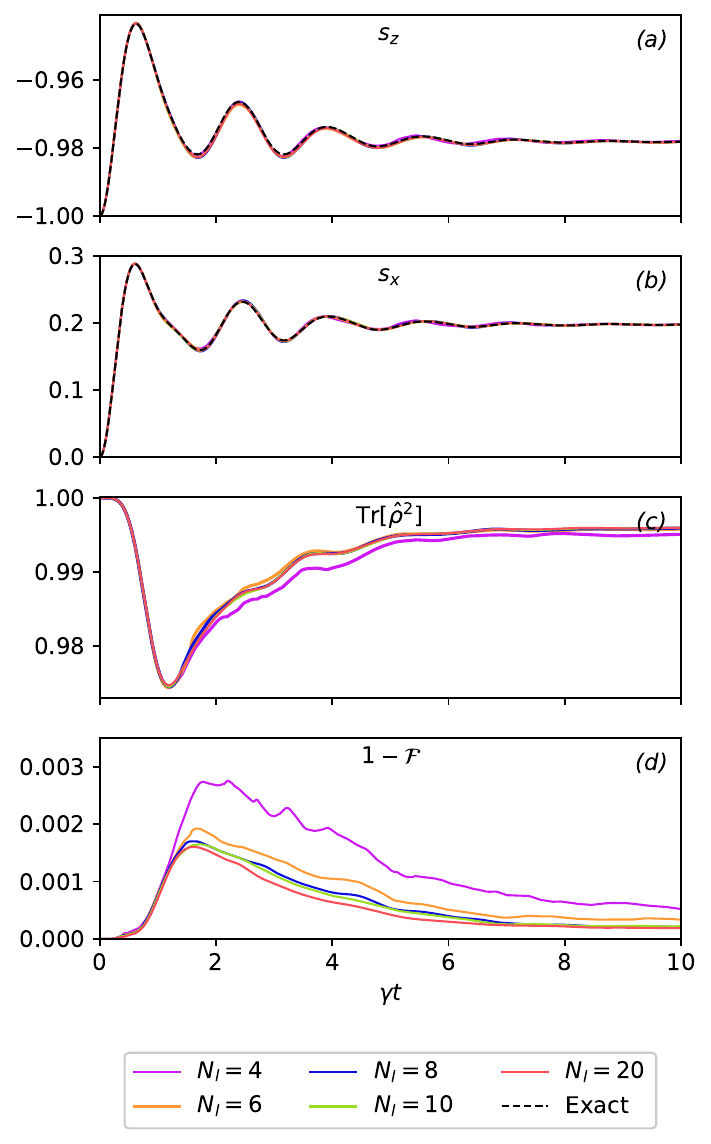}
    \caption{Time evolution of a $3\times3$ dissipative TFIM with $J_z=\gamma=1$, $h=0.5$ and $dt=5\times 10^{-3}$ computed with ansatz II with rank $R=10$: (a) average magnetization along the $x$-axis $s_x$, (b) average magnetization along the $z$-axis $s_z$, (c) purity of the variational state $\hat{\rho}$, (d) infidelity $1-\mathcal{F}$ between the variational and exact quantum states.}
    \label{Fig.SV_ansatz2}
    \end{figure}
Overall, both ansatze achieve very a high fidelity, highlighting the effectiveness of the algorithm. In particular, both schemes excel at capturing the final steady-state. Notably, the infidelity displays a maximum around $\gamma t = 2$, which coincides with a minimum in the state purity. For ansatz I, the peak infidelity decreases when increasing the rank $R$ (see Fig.~\ref{Fig.SV_ansatz1_rank}), but not when increasing the number of layers (see Fig.~\ref{Fig.SV_ansatz1}(d)). This result indicates that ansatz I is less effective in representing faithfully all the distinct states occurring in the statistical mixture. Conversely, increasing the number of layers of ansatz II is sufficient to decrease the overall infidelity. 
Finally, we compare the integrated Bures distance~\cite{Laha_2021} to the exact solution, given by
\begin{equation}
    I_{B}(T) = \frac{1}{T}\int_0^T \sqrt{2\textrm{Tr}\left[\sqrt{\hat{\rho}}\hat{\rho}_0\sqrt{\hat{\rho}}\right]}dt\,,
\end{equation}
where, $\hat{\rho}_0$ is the exact state of the system, computed with QuTiP, at $\gamma t = 7$. The results are presented in Fig.~\ref{Fig.BE}, which illustrates the trade-off between the computational resources and accuracy. As expected, ansatz II achieves the same accuracy as ansatz I with fewer layers. However, it requires a considerably larger number of parameters and circuit measurements to achieve the same target accuracy. Ultimately, the choice between the two ansatze depends on the problem at hand. Specifically, whether we need to optimize time and memory requirements, or rather keep a low circuit depth to minimize quantum error. In Appendix \ref{Cost}, a derivation of the computational cost of both ansatze is presented, in terms of the number of distinct quantum circuits required to estimate the entries of $M$ and $V$.
\begin{figure}[h!]
    \centering
        \centering
        \includegraphics[width = 0.5\textwidth]{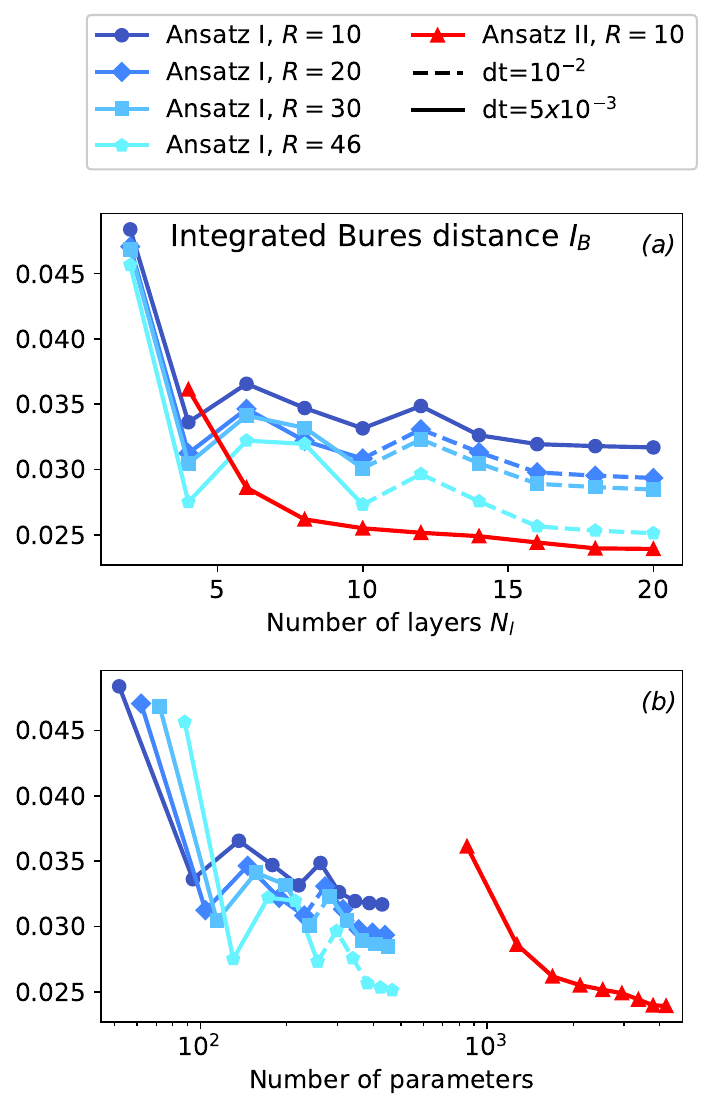}
        \label{Fig.SV2}
    \caption{The accuracy measured by the integrated Bures error $I_{B}$ at $\gamma t = 7$ as a function of (a) the number of layers and (b) the total number of angular variational parameters for ansatze I and II.}
    \label{Fig.BE}
\end{figure}
\subsection{Simulation with noise}
\begin{figure}[t!]
   \centering
    \includegraphics[width = 0.5\textwidth]{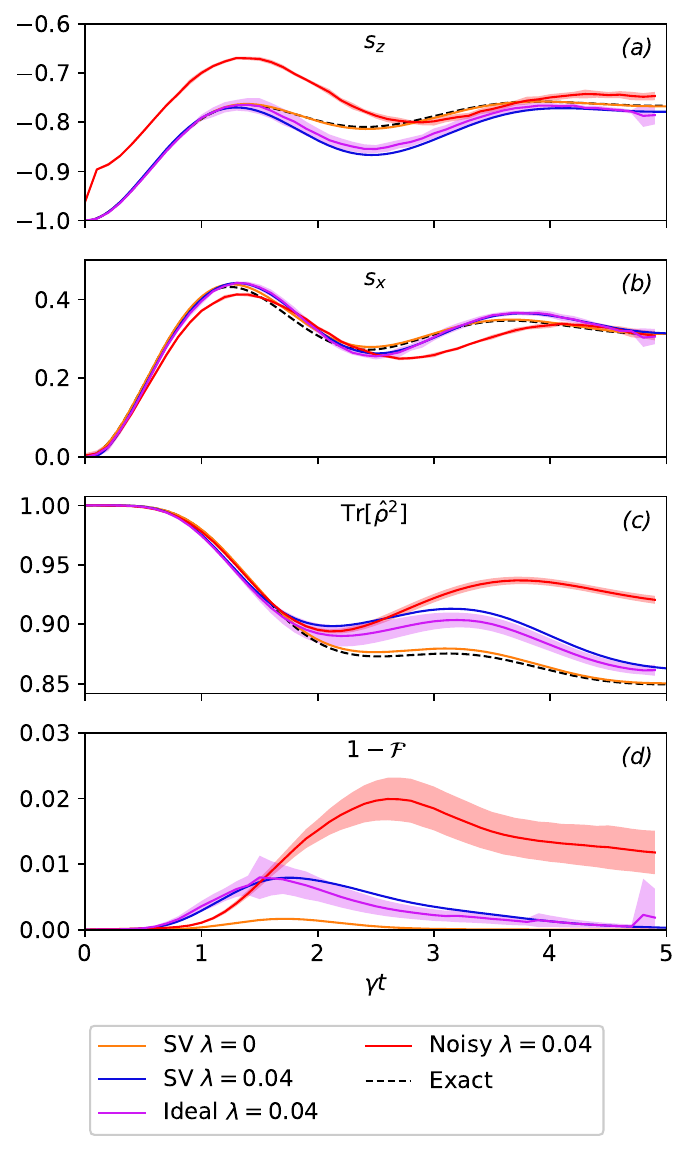}
       \caption{Time evolution of a 2-site dissipative TFIM with $J_z=\gamma=1$, $h=0.5$ and $dt=1\times 10^{-2}$ computed with ansatz I with rank $R=4$: (a) average magnetization along the $x$-axis $s_x$, (b) average magnetization along the $z$-axis $s_z$, (c) purity of the variational state $\hat{\rho}$, (d) infidelity $1-\mathcal{F}$ between the variational and exact quantum states. Each curve was obtained with different quantum simulators: 'Exact' - QuTiP, 'SV' - statevector, 'Ideal' - IBM's Qasm Simulator with no hardware noise, and 'Noisy' - Qasm Simulator, with the noise model of IBM's Lagos machine. For the 'Ideal' and 'Noisy' curves, each point in time is an average of 10 simulations of the corresponding time step and the shaded area is their standard deviation. For Qiskit's simulations, each quantum circuit is run 20000 times.}
    \label{Fig.6}
\end{figure}

\begin{figure}[t!]
    \centering
    \includegraphics[width = 0.48\textwidth]
    {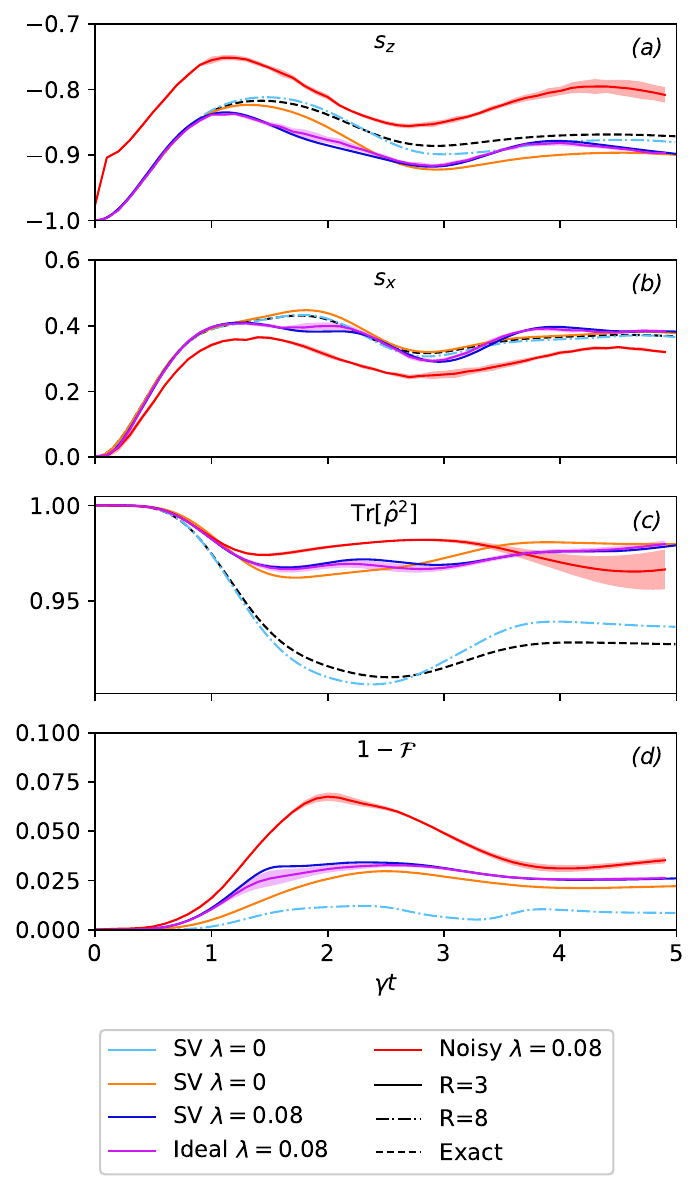}
    \caption{Time evolution of a 3-site dissipative TFIM $J_z=\gamma=1$, $h=0.5$ and $dt=1\times 10^{-2}$ computed with ansatz I with rank $R=3$ (solid lines) and $8$ (dash-dotted lines): (a) average magnetization along the $x$-axis $s_x$, (b) average magnetization along the $z$-axis $s_z$, (c) purity of the variational state $\hat{\rho}$, (d) infidelity $1-\mathcal{F}$ between the variational and exact quantum states. For the 'Ideal' and 'Noisy' curves, each point in time is an average of 10 simulations of the corresponding time step and the shaded area is the standard deviation. For the Qiskit simulations, each quantum circuit is run 20000 times.}
    \label{Fig.7}
\end{figure}

We simulate a 2-site and a 3-site (1D) dissipative TFIM model using IBM's Qasm Simulator~\cite{gadi_aleksandrowicz_2019_2562111}. We restrict this analysis to ansatz I and $N_{l}=2$, which represents faithfully the state of such small systems. Results are displayed in Fig.~\ref{Fig.6}. 
We compare results obtained using a noise model corresponding to the IBM Lagos quantum device, with noiseless simulations (i.e., including only the shot noise from projective readout), and with statevector simulations. Noisy estimates of the matrix elements can lead to a non semi-positive defined matrix $M$. As the matrix is typically singular and must be regularized to solve the system Eq.~\eqref{eq:MV_MA}, the occurrence of negative eigenvalues calls for a different regularization scheme, as detailed in Appendix \ref{regularization}. For details on the estimation of $M$ and $V$, we refer the reader to Appendix \ref{QCircuits}.
Fig.\ref{Fig.7} presents the results of the full-rank QTE for the 2-site dissipative TFIM. We compare the results of the statevector, noiseless, and noisy simulations of the $x$- and $z$-polarizations, and we study the purity and the fidelity to the exact results. To compute the purity and the fidelity in the noiseless and noisy simulations, a full estimate of the density matrix at each time would be needed. Instead of simulating a computationally demanding full-state tomography, we resort to a hybrid approach consisting in computing the variational parameters from the noiseless and noisy simulations, and using them to compute the density matrix directly from Eq.~\eqref{eq:a1}.
The results for the average polarizations indicate that the noiseless simulation reproduces accurately the statevector simulation, when using the same regularization ($\lambda=0.04$, see Appendix~\ref{regularization} for details about the regularization parameter $\lambda$). However, the noisy simulation shows a significant bias, which is due to both the limited accuracy in estimating $M$ and $V$, and the error in the estimation of the observables. This analysis is further supported by the study of the purity and infidelity.
We also demonstrate the algorithm on a 3-site dissipative TFIM with open boundary conditions, taking $R=3$ and selecting the computational basis vectors $\{\ket{000},\ket{001},\ket{010}\}$ (Fig.~\ref{Fig.7}). The statevector simulation deviates from the exact result due to the low-rank approximation, as confirmed by the statevector simulation with $R=8$. As in the previous case, the noise leads to a significant bias for the observables. The low-rank approximation also affects the purity and fidelity, which now stay finite in the long-time limit.

\section{Conclusions \& Outlook} \label{conclusion}
In this paper, we propose a novel time-dependent variational quantum algorithm, which directly integrates the Lindblad master equation. In this framework, the state of the system is approximated by a low-rank ansatz, defined as a linear combination of pure states that are represented by parameterized quantum circuits. The time-evolution of the system is then achieved by evolving the variational parameters, according to McLachlan's variational principle. This algorithm requires only evaluating circuits with $n$ qubits, less than other existing similar algorithms.

As a demonstration of the algorithm and a proof of concept, we applied it to a dissipative transverse field Ising system. Two types of simulations were performed, one based on the measurement of quantum circuits with Qiskit's Qasm Simulator, with and without hardware noise, and a noise-free statevector simulation, which allowed us to verify the performance of the algorithm when simulating larger systems and study the expressivity error of the variational ansatze.

The statevector simulations were implemented with two different ansatze, respectively expressing a statistical mixture of orthogonal or non-orthogonal pure states. We find that the orthogonal ansatz is more resource-efficient. Nevertheless, the non-orthogonal scheme has a more flexible structure and can achieve better accuracy with shallower quantum circuits.

Statevector simulations showed that the algorithm can efficiently model the open system dynamics in cases where the system entropy remains bounded, at a computational cost which scales only linearly with the rank. 

A considerable advantage of the present algorithm is that it requires only $n$ qubits, i.e., as many as are needed to encode the pure states $\ket{\psi_i}$. This comes as a considerable advantage compared to most algorithms relying on purification of the quantum state \cite{Yuan2019theoryofvariational} or vectorization of the density matrix \cite{PhysRevResearch.2.043289}, which instead require $2n$ qubits. This advantage is particularly relevant in view of the application of error mitigation strategies \cite{cai2023quantum,Mari_2021,berg2022probabilistic,Endo_2018,Endo_2021,Nation_2021,Yang_2022} whose effectiveness depends critically both on the width and depth of a quantum circuit.

Simulations accounting for noise on Qiskit’s Qasm simulator showed that accurate regularization of the matrix $M$ defining the variational dynamics is needed in order to avoid numerical instabilities. Here, we applied the diagonal shift regularization method, which preserves the positivity of $M$. Noise may additionally lead to biases in the estimate of expectation values, hence to inaccurate evolution of the variational parameters. This issue can be at least partially addressed by quantum error mitigation techniques, which will demand further investigation.

A possible generalization of the present algorithm is the implementation of an adaptive-rank scheme, similarly to what has been proposed for classical simulations of open quantum dynamics \cite{gravina2023adaptive,Donatella_2021,chen2020low,LeBris_2013_low_rank,McCaul_2021}. An adaptive-rank scheme would avoid accounting for irrelevant states, i.e., with vanishing probabilities, in the statistical mixture, thereby increasing computational efficiency while reducing numerical instabilities due to small probabilities. An adaptive-rank scheme may also be used in cases where the transient dynamics is not necessarily low-rank while the long-time limit is. In these cases, an adaptive bounded-rank simulation of a fictitious system dynamics would efficiently describe the long-time limit.

Another possible improvement is the generalization of the projected variational quantum dynamics to the present low-rank algorithm. For unitary quantum dynamics, the p-VQD approach was successful in addressing the occurrence of biases and uncontrolled variance in Monte-Carlo estimates, arising from vanishing amplitudes in the variational wave functions \cite{Barison2021efficientquantum,GaconPRR2024,Sinibaldi_2023}. We expect a similar improvement in the present case.

\section{Acknowledgements}
Sara Santos and Xinyu Song contributed equally to this work. We thank Zakari Denis for insightful discussions. We acknowledge the use of IBM Quantum services for this work. The views expressed are those of the authors,
and do not reflect the official policy or position of IBM
or the IBM Quantum team. This work was supported as a part of NCCR SPIN, a National Centre of Competence in Research, funded by the Swiss National Science Foundation (grant number 225153).

\bibliographystyle{quantum}
\bibliography{bibliography}

\onecolumn
\pagebreak
\appendix
\section{McLachlan's variational principle}
\label{MCPrinciple}
\subsection{Derivation in the density matrix formalism}
\par The starting point of our derivation is the variational density matrix $\hat{\rho}(\bm{\beta})$ and the Lindblad master equation $\dot{\hat{\rho}} = \mathcal{L}[\hat{\rho}]$. Following the McLachlan's variational principle~\cite{Yuan2019theoryofvariational}, the time evolution of the variational parameters is obtained from minimizing the $L^2$-distance between the variational state $\hat{\rho}(\bm{\beta}(t)) + \sum_j \frac{\partial \hat{\rho}(\bm{\beta})}{\partial \beta_j}\dot{\beta_j}\delta t$ and the exact time evolved state $\hat{\rho}(\bm{\beta}(t)) + \mathcal{L}\left[\hat{\rho}(\bm{\beta}(t))\right] \delta t$. This variational principle can be expressed as follows:
\begin{equation}
     \boldsymbol{\dot{\beta}} = \textrm{argmin}\left(\lVert \sum_j \frac{\partial \hat{\rho}(\bm{\beta})}{\partial \beta_j}\dot{\beta_j}-\mathcal{L}\left[\hat{\rho}(\bm{\beta}(t))\right] \rVert_2\right) \,,
\end{equation}
or equivalently,
\begin{equation}
        \boldsymbol{\dot{\beta}} = \textrm{argmin} \left( C(\boldsymbol{\dot{\beta}}) \right) \, ,
\end{equation}
\begin{equation}{\label{eq:CostF}}
C(\boldsymbol{\dot{\beta}}) = \mathrm{Tr}\left[\left(\sum_j \frac{\partial \hat{\rho}(\bm{\beta})}{\partial \beta_j}\dot{\beta_j}-\mathcal{L}\left[\hat{\rho}(\bm{\beta}(t))\right]\right)^{\dagger} 
\left(\sum_j \frac{\partial \hat{\rho}(\bm{\beta})}{\partial \beta_j}\dot{\beta_j}-\mathcal{L}\left[\hat{\rho}(\bm{\beta}(t))\right]\right)\right] \,,
\end{equation}
where $C(\bm{\dot{\beta}})$ can be interpreted as a cost function. To find the minimum of $C(\bm{\dot{\beta}})$, we calculate the derivative of $C(\bm{\dot{\beta}})$ with respect to the variable $\dot{\beta_k}$ 
\begin{equation}
\begin{aligned}
\begin{split}
    \frac{\partial C(\bm{\dot{\beta}})}{\partial \dot{\beta_k}} &= \mathrm{Tr}\left[\frac{\partial \hat{\rho}^{\dagger}}{\partial \beta_k}\left(\sum_j \frac{\partial \hat{\rho}}{\partial \beta_j}\dot{\beta_j} -  \mathcal{L}[\hat{\rho}]\right)+\left(\sum_j \frac{\partial \hat{\rho}}{\partial \beta_j}\dot{\beta_j} - \mathcal{L}[\hat{\rho}]\right)^{\dagger}\frac{\partial\hat{\rho}}{\partial \beta_k}\right]\\
    & = \sum_j \left(\mathrm{Tr}\left[\frac{\partial \hat{\rho}^{\dagger}}{\partial {\beta_k}}\frac{\partial \hat{\rho}}{\partial \beta_j}\right]\dot{\beta_j}+h.c.\right)
    - \left(\mathrm{Tr}\left[\frac{\partial \hat{\rho}^{\dagger}}{\partial \beta_k}\mathcal{L}[\hat{\rho}]\right]+h.c.\right)\\
    &= 2\left(\sum_j\mathcal{R}e\left[\mathrm{Tr}\left[\frac{\partial \hat{\rho}}{\partial {\beta_k}}\frac{\partial \hat{\rho}}{\partial \beta_j}\right]\right]\dot{\beta_j}-\mathcal{R}e\left[\mathrm{Tr}\left[\frac{\partial \hat{\rho}}{\partial {\beta_k}}\mathcal{L}[\hat{\rho}]\right]\right]\right) \\
    & = 2\left(\sum_j\mathrm{Tr}\left[\frac{\partial \hat{\rho}}{\partial {\beta_k}}\frac{\partial \hat{\rho}}{\partial \beta_j}\right]\dot{\beta_j}- \mathrm{Tr}\left[\frac{\partial \hat{\rho}}{\partial {\beta_k}}\mathcal{L}[\hat{\rho}]\right]\right) \,,
\end{split}
\end{aligned}
\end{equation}
where we have used the hermiticity of the density matrix. By setting $\frac{\partial C(\bm{\dot{\beta}})}{\partial \dot{\beta_k}} = 0$, we obtain the equations of motion (EOM) of the variational parameters, presented in Eq.~\eqref{eq:MV_MA}.
\subsection{Equations of motion - ansatz I}
\label{EOMI}
    \par In this section we derive the EOM of the parameters $\bm{\beta}=(\bm{\alpha},\bm{\theta})$ by computing the elements $M_{kj}$ and $V_k$ according to Eq.~\eqref{eq:MV_MA}. Let $\hat{\rho}(\bm{\alpha},\bm{\theta}) = \sum_{p=1}^R \alpha_p \hat{U}(\bm{\theta})\ket{\mathbf{x}_p}\bra{\mathbf{x}_p} \hat{U}(\bm{\theta})^{\dagger}$ be the density matrix and $\ket{\mathbf{x}_p}$ computational basis states. For simplicity, we use the notation $\hat{U}(\bm{\theta}) = \hat{U}$. We begin by writing the derivatives:
    \begin{equation}\label{eq.deriv_alpha}
        \frac{\partial\hat{\rho}}{\partial\alpha_p} = \hat{U}\ket{\mathbf{x}_p}\bra{\mathbf{x}_p}\hat{U}^{\dagger}\,,
    \end{equation}
    and
    \begin{equation}\label{eq.deriv_theta}
        \frac{\partial\hat{\rho}}{\partial\theta_j} = \sum_p \alpha_p\left(\frac{\partial \hat{U}}{\partial \theta_j}\ket{\mathbf{x}_p}\bra{\mathbf{x}_p}\hat{U}^{
        \dagger} +\hat{U}\ket{\mathbf{x}_p}\bra{\mathbf{x}_p}\frac{\partial \hat{U}^{\dagger}}{\partial \theta_j} \right) \,.
    \end{equation}
    Plugging Eqs.\eqref{eq.deriv_alpha} and \eqref{eq.deriv_theta} in Eq.\eqref{eq:MV_MA}, we obtain the entries of the $M$. First we compute $M_{\alpha_p\alpha_q}$, corresponding to the $\alpha_p$ and $\alpha_q$ parameters:

    \begin{equation} \label{eq:Malal1}
    \begin{split}
            M_{\alpha_p\alpha_q}&=\mathrm{Tr}\left[{\frac{\partial \hat{\rho}^{\dagger}}{\partial \alpha_p}\frac{\partial \hat{\rho}}{\partial \alpha_q}}\right]\\ &=\mathrm{Tr}\left[\hat{U}\ket{\mathbf{x}_p}\bra{\mathbf{x}_p}\hat{U}^{\dagger}\hat{U}\ket{\mathbf{x}_q}\bra{\mathbf{x}_q}\hat{U}^{\dagger}\right]\\&=\mathrm{Tr}\left[\hat{U}\ket{\mathbf{x}_p}\delta_{p,q}\bra{\mathbf{x}_q}\hat{U}^{\dagger}\right]\\&=\mathrm{Tr}\left[\ket{\mathbf{x}_p}\bra{\mathbf{x}_q}\hat{U}^{\dagger}\hat{U}\right]\delta_{p,q}\\
            &= \delta_{p,q}\delta_{p,q} = \delta_{p,q}\,,
    \end{split}
    \end{equation}
    where $\delta_{p,q}$ is the Kronecker delta. The remaining matrix elements $M_{\alpha_p\theta_j}$ and $M_{\theta_k\theta_j}$ are then given by

    \begin{equation} \label{eq:Malth1}
    \begin{split}
            M_{\alpha_p\theta_j}&= \mathrm{Tr}\left[\frac{\partial \hat{\rho}^{\dagger}}{\partial \alpha_p}\frac{\partial \hat{\rho}}{\partial \theta_j}\right]\\ 
            &=\mathrm{Tr}\left[\hat{U}\ket{\mathbf{x}_p}\bra{\mathbf{x}_p}\hat{U}^{\dagger} \sum_q \alpha_q\left(\frac{\partial \hat{U}}{\partial \theta_j}\ket{\mathbf{x}_q}\bra{\mathbf{x}_q}\hat{U}^{\dagger} +\hat{U}\ket{\mathbf{x}_q}\bra{\mathbf{x}_q}\frac{\partial \hat{U}^{\dagger}}{\partial\theta_j}\right)\right]\\
            &= \sum_q \alpha_q \delta_{p,q}\mathrm{Tr}\left[\frac{\partial \hat{U}}{\partial \theta_j}\ket{\mathbf{x}_q}\bra{\mathbf{x}_p}\hat{U}^{\dagger}+\hat{U}\ket{\mathbf{x}_p}\bra{\mathbf{x}_q}\frac{\partial \hat{U}^{\dagger}}{\partial \theta_j}\right] \\
            &=\alpha_p\mathrm{Tr}\left[\ket{\mathbf{x}_p}\bra{\mathbf{x}_p}\hat{U}^{\dagger}\frac{\partial \hat{U}}{\partial \theta_j}+\ket{\mathbf{x}_p}\bra{\mathbf{x}_p}\frac{\partial \hat{U}^{\dagger}}{\partial\theta_j}\hat{U}\right]\\
            &=\alpha_p\mathrm{Tr}\left[\ket{\mathbf{x}_p}\bra{\mathbf{x}_p}\left(\hat{U}^{\dagger}\frac{\partial \hat{U}}{\partial \theta_j}+\frac{\partial \hat{U}^{\dagger}}{\partial\theta_j}\hat{U}\right)\right]\\
            &=\alpha_p\mathrm{Tr}\left[\ket{\mathbf{x}_p}\bra{\mathbf{x}_p}\frac{\partial \hat{U}^{\dagger}\hat{U}}{\partial \theta_j}\right]\\           
            &=\alpha_p\mathrm{Tr}\left[\ket{\mathbf{x}_p}\bra{\mathbf{x}_p}\frac{\partial \mathbb{I}}{\partial \theta_j}\right]=0 \, ,
    \end{split}
    \end{equation}
    and
    \begin{equation} \label{eq:Mthth1}
    \begin{split}
            M_{\theta_k\theta_j} &= \mathrm{Tr}\left[\frac{\partial \hat{\rho}^{\dagger}}{\partial \theta_k}\frac{\partial \hat{\rho}}{\partial \theta_j}\right] \\
            &=\mathrm{Tr}\left[\sum_p\alpha_p\left(\frac{\partial \hat{U}}{\partial \theta_k}\ket{\mathbf{x}_p}\bra{\mathbf{x}_p}\hat{U}^{\dagger} +\hat{U}\ket{\mathbf{x}_p}\bra{\mathbf{x}_p}\frac{\partial\hat{U}^{\dagger}}{\partial \theta_k}\right) \sum_q\alpha_q\left(\frac{\partial \hat{U}}{\partial \theta_j}\ket{\mathbf{x}_q}\bra{\mathbf{x}_q}\hat{U}^{\dagger} +\hat{U}\ket{\mathbf{x}_q}\bra{\mathbf{x}_q}\frac{\partial \hat{U}^{\dagger}}{\partial \theta_j} \right)\right]\\
            &=\mathrm{Tr}\left[\sum_{p,q}\alpha_p\alpha_q \left(\frac{\partial \hat{U}}{\partial \theta_k}\ket{\mathbf{x}_p}\bra{\mathbf{x}_p}\hat{U}^{\dagger}\frac{\partial \hat{U}}{\partial \theta_j}\ket{\mathbf{x}_q}{\bra{\mathbf{x}_q}}\hat{U}^{\dagger}+\frac{\partial \hat{U}}{\partial \theta_k}\ket{\mathbf{x}_p}\bra{\mathbf{x}_q}\frac{\partial \hat{U}^{\dagger}}{\partial \theta_j}\delta_{p,q}+\delta_{p,q}\braket{\mathbf{x}_p|\frac{\partial \hat{U}^{\dagger}}{\partial \theta_k}\frac{\partial \hat{U}}{\partial \theta_j}|\mathbf{x}_q} \right.\right.\\
            &+\left.\left. \hat{U}\ket{\mathbf{x}_p}\bra{\mathbf{x}_p}\frac{\partial \hat{U}^{\dagger}}{\partial \theta_k}\hat{U}\ket{\mathbf{x}_q}\bra{\mathbf{x}_q}\frac{\partial \hat{U}^{\dagger}}{\partial \theta_j} \right) \right]\\
            &=2\mathcal{R}e\left[\sum_p\alpha_p^2\bra{\mathbf{x}_p}\frac{\partial \hat{U}^{\dagger}}{\partial \theta_k}\frac{\partial \hat{U}}{\partial \theta_j}\ket{\mathbf{x}_p}+\sum_{p,q}\alpha_p\alpha_q\bra{\mathbf{x}_p}\frac{\partial \hat{U}^{\dagger}}{\partial \theta_k}\hat{U}\ket{\mathbf{x}_q}\bra{\mathbf{x}_q}\frac{\partial \hat{U}^{\dagger}}{\partial \theta_j}\hat{U}\ket{\mathbf{x}_p}\right]  \,.        
    \end{split}
    \end{equation}
The Liouvillian operator can be expanded, w.l.o.g. as $\mathcal{L}=\sum_r c_r \hat{\sigma}_r^1\hat{\rho} \hat{\sigma}_r^2$, with coefficients $c_r$ and Pauli strings $\hat{\sigma}_r^{1}$, $\hat{\sigma}_r^{2}$. The components of $V$ can be then expressed as
    \begin{equation} \label{eq:Val1}
        \begin{split}
            V_{\alpha_p} &= \mathrm{Tr}\left[\frac{\partial \hat{\rho}}{\partial \alpha_p}\mathcal{L}\left[\hat{\rho}\right]\right]\\ &=\mathcal{R}e\left[\sum_{r,q}c_r\alpha_q\bra{\mathbf{x}_p}\hat{U}^{\dagger}\hat{\sigma}_r^1\hat{U}\ket{\mathbf{x}_q}\bra{\mathbf{x}_q}\hat{U}^{\dagger}\hat{\sigma}_r^{2}\hat{U}\ket{\mathbf{x}_p}\right] \,,
        \end{split}
    \end{equation}
    and
    \begin{equation} \label{eq:Vth1}
        \begin{split}
            V_{\theta_k} &=\mathrm{Tr}\left[\frac{\partial \hat{\rho}}{\partial \theta_k}\mathcal{L}\left[\hat{\rho}\right]\right]\\ &=\mathcal{R}e\left[\sum_{rpq}c_r\alpha_p\alpha_q\left(\bra{\mathbf{x}_p}\frac{\partial \hat{U}^{\dagger}}{\partial \theta_k}\hat{\sigma}_r^1\hat{U}\ket{\mathbf{x}_q}\bra{\mathbf{x}_q}\hat{U}^{\dagger}\hat{\sigma}_r^{2}\hat{U}\ket{\mathbf{x}_p}
            +\bra{\mathbf{x}_p}\frac{\partial \hat{U}^{\dagger}}{\partial \theta_k}\hat{\sigma}_r^{2}\hat{U}\ket{\mathbf{x}_q}^* \bra{\mathbf{x}_p}\hat{U}^{\dagger}\hat{\sigma}_r^1\hat{U}\ket{\mathbf{x}_q}\right)\right]\,.
        \end{split}
    \end{equation}
\subsection{Equations of motion - ansatz II}
\label{EOMII}
    \par Proceeding similarly to section \ref{EOMI}, we derive the EOM of the parameters $\bm{\beta}=(\bm{\alpha},\bm{\theta})$ by computing the elements $M_{kj}$ and $V_k$ for ansatz II $\hat{\rho}(\bm{\beta)} = \sum_{p=1}^{R} \alpha_p \hat{U}(\bm{\theta}^{(p)})\ket{\mathbf{x}_p}\bra{\mathbf{x}_p} \hat{U}(\bm{\theta}^{(p)})^{\dagger}$, with $\bm{\theta}=(\bm{\theta}^{(0)},\bm{\theta}^{(1)},...)$ be the set of variational angular parameters. Again, we take for simplicity $\hat{U}(\bm{\theta}^{(p)}) = \hat{U}_p$. We begin by writing explicitly the derivatives
    \begin{equation}
        \frac{\partial\hat{\rho}}{\partial\alpha_p} = \hat{U}_p\ket{\mathbf{x}_p}\bra{\mathbf{x}_p}\hat{U}_p^{\dagger}\,,
    \end{equation}
    \begin{equation}
    \begin{split}
        \frac{\partial\hat{\rho}}{\partial\theta_j^{(q)}} = \alpha_q \left(\frac{\partial \hat{U}_q}{\partial \theta_j^{(q)}}\ket{\mathbf{x}_q}\bra{\mathbf{x}_q}\hat{U}_q^{\dagger}
        +\hat{U}_q\ket{\mathbf{x}_q}\bra{\mathbf{x}_q}\frac{\partial \hat{U}_q^{\dagger}}{\partial \theta_j^{(q)}}\right)\,.
    \end{split}
    \end{equation}
    Plugging these in Eq.\eqref{eq:MV_MA}, we obtain the elements $M_{\alpha_p\alpha_q}$ corresponding to the $\alpha_p$ and $\alpha_q$ parameters:
    \begin{equation} \label{eq:Malal2}
    \begin{split} M_{\alpha_p\alpha_q}&=|\bra{\mathbf{x}_p}\hat{U}_p^{\dagger}\hat{U}_q\ket{\mathbf{x}_q}|^2\,.
    \end{split}
    \end{equation}
    The remaining matrix elements $M_{\alpha_p\theta_j^{(q)}}$ and $M_{\theta_k^{(p)}\theta_j^{(q)}}$ are simply
    \begin{equation} \label{eq:Malth2}
    \begin{split}    M_{\alpha_p\theta_j^{(q)}}&=2\alpha_p \mathcal{R}e\left[\bra{\mathbf{x}_p}\hat{U}_p^{\dagger}\frac{\partial \hat{U}_q}{\partial \theta_j^{(q)}}\ket{\mathbf{x}_q}\bra{\mathbf{x}_q}\hat{U}_q^{\dagger}\hat{U}_p\ket{\mathbf{x}_p}\right]\,,
    \end{split}
    \end{equation}
    and
    \begin{equation} \label{eq:Mthth2}
    \begin{split}
            M_{\theta_k^{(p)}\theta_j^{(q)}} &= 2\alpha_p\alpha_q \mathcal{R}e\left[\bra{\mathbf{x}_p}\hat{U}_p^{\dagger}\hat{U}_q\ket{\mathbf{x}_q}
            \bra{\mathbf{x}_q}\frac{\partial \hat{U}_q^{\dagger}}{\partial \theta_j^{(q)}}\frac{\partial \hat{U}_p}{\partial \theta_k^{(p)}}\ket{\mathbf{x}_p} \right.\\
            &+\left.\bra{\mathbf{x}_p}\hat{U}_p^{\dagger}\frac{\partial \hat{U}_q}{\partial \theta_j^{(q)}}\ket{\mathbf{x}_q}\bra{\mathbf{x}_q}\hat{U}_q^{\dagger}\frac{\partial \hat{U}_p}{\partial \theta_k^{(p)}}\ket{\mathbf{x}_p}\right] \,.
    \end{split}
    \end{equation}
    Similarly, expanding the Liouvillian operator as $\mathcal{L}[\hat{\rho}]=\sum_r c_r \hat{\sigma_r}^1\hat{\rho} \hat{\sigma_r}^2$, the elements of the $V$ vector are given by
    \begin{equation} \label{eq:Val2}
        \begin{split}
            V_{\alpha_p} = \mathcal{R}e\left[\sum_{r,q}c_r\alpha_q\bra{\mathbf{x}_p}\hat{U}_p^{\dagger}\hat{\sigma}_r^1\hat{U}_q\ket{\mathbf{x}_q}           
            \bra{\mathbf{x}_q}\hat{U}_q^{\dagger}\hat{\sigma}_r^{2}\hat{U}_p\ket{\mathbf{x}_p}\right] \,,
        \end{split}
    \end{equation}
    and
    \begin{equation} \label{eq:Vth2}
        \begin{split}
            V_{\theta_k^{(p)}}&=\alpha_p \mathcal{R}e\left[\sum_{rq}c_r\alpha_q\left(\bra{\mathbf{x}_p}\frac{\partial \hat{U}_p^{\dagger}}{\partial \theta_k^{(p)}}\hat{\sigma}_r^1\hat{U}_q\ket{\mathbf{x}_q} 
            \times\bra{\mathbf{x}_q}\hat{U}_q^{\dagger}\hat{\sigma}_r^{2}\hat{U}_p\ket{\mathbf{x}_p}\right.\right.  \\
            &+\left.\left.\bra{\mathbf{x}_p}\frac{\partial \hat{U}_p^{\dagger}}{\partial \theta_k^{(p)}}\hat{\sigma}_r^{2}\hat{U}_q\ket{\mathbf{x}_q}^* \bra{\mathbf{x}_p}\hat{U}_p^{\dagger}\hat{\sigma}_r^1\hat{U}_q\ket{\mathbf{x}_q}\right)\right] \,.
        \end{split}
    \end{equation}

\section{Truncated basis}
\label{basis}
\begin{figure}[h!]
\centering
\includegraphics[width = 0.6\textwidth]{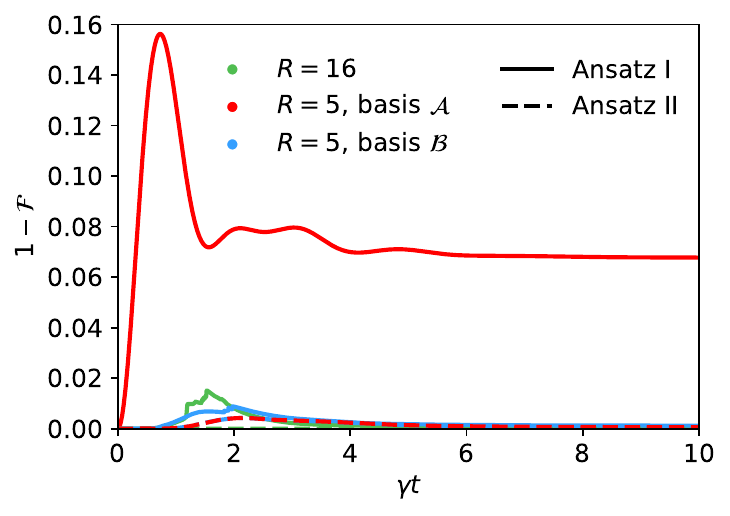}
\caption{Influence of the initial basis choice on the performance of the LRQTE algorithm for ansatze I and II.}
\label{Fig.basis_comp}
\end{figure}

When implementing LRQTE with a truncated basis, one must choose among the $2^n$ possible basis states. In this work, we made this selection based on the expansion of the Kraus operators $\hat{K}_k$ that govern the time dynamics of the system. Let $\hat{\rho}$ be the density matrix of an open quantum system. The time evolved state, given by the Lindblad master equation Eq.~\eqref{eq:1}, can be expressed as a Kraus map~\cite{McCaul_2021}
\begin{equation}
\label{kraus_exp}
\hat{\rho}(t + \delta t) = \frac{1}{2K} \sum_{k=1}^K \left( \mathcal{\hat{U}}_k \hat{\rho}(t) \mathcal{\hat{U}}_k^\dagger + \mathcal{\hat{V}}_k \hat{\rho}(t) \mathcal{\hat{V}}_k^\dagger\right) + \mathcal{O}(\delta t^2) \,,
\end{equation}
where $\{\mathcal{\hat{U}}_k\}_{k=1}^K$ and $\{\mathcal{\hat{V}}_k\}_{k=1}^K$ are the Kraus operators of the dynamics defined as
\begin{equation}
    \mathcal{\hat{U}}_k = e^{\delta t \hat{J}_k -i \sqrt{\gamma \delta t} \hat{c}_k}\,,
\end{equation}
\begin{equation}
    \mathcal{\hat{V}}_k = e^{\delta t \hat{J}_k +i \sqrt{\gamma \delta t} \hat{c}_k}\,,
\end{equation}
\noindent with $\hat{J}_k = -i\hat{H} -\frac{\gamma}{2}\left(\hat{c}_k^2 - \hat{c}_k^\dagger \hat{c}_k\right)$. Using the Zassenhaus approximation~\cite{zassenhaus}, we obtain
\begin{equation}
\label{eq:zass}
    \mathcal{\hat{U}}_k \hat{\rho}(t) \mathcal{\hat{U}}_k^\dagger + \mathcal{\hat{V}}_k \hat{\rho}(t) \mathcal{\hat{V}}_k^\dagger = e^{\delta t\hat{J}_k}\hat{\Gamma}_k e^{\delta t\hat{J}_k^\dagger} + \mathcal{O}(\delta t^2)\,,
\end{equation}
\noindent where $\hat{\Gamma}_k = \hat{\rho}(t) + \gamma \delta t \left(\hat{c}_k\hat{\rho}(t)\hat{c}_k^\dagger -\hat{c}_k^2\hat{\rho}(t)\right) + h.c.$. For the dissipative TFIM, with jump operators $\hat{c}_k = \hat{\sigma}_k^{-}$, $\hat{c}_k^2  = 0$, we can further simplify $\hat{J}_k = -i\hat{H} +\frac{K}{2} \hat{\sigma}_k^{+}\hat{\sigma}_k^{-}$ and $\hat{\Gamma}_k = \hat{\rho}(t) +\gamma \delta t \hat{\sigma}_k^-\hat{\rho}(t)\hat{\sigma}_k^+ + h.c.$. Consider the initial state $\hat{\rho}(0) = |1\rangle\langle 1|_n$. According to Eqs.~\eqref{kraus_exp} and~\eqref{eq:zass}, the density matrix at a later time $\delta t$ is thus simply given by
\begin{equation}
    \hat{\rho}(\delta t) = \frac{1}{K} \sum_{k=1}^K \left(e^{-i\delta t\hat{H}}\hat{\rho}(0) e^{i\delta t\hat{H}^\dagger} \right) + \mathcal{O}(\delta t^2)\,,
    \label{eq.time_rho}
\end{equation}
where we have used $\hat{\sigma}_k^{-}|1\rangle\langle 1|_n\hat{\sigma}_k^{+}=0$. Eq.~\eqref{eq.time_rho} describes the Hamiltonian dynamics of a quantum system isolated from its environment.
In this picture, the time-evolved state of the system is approximately
\begin{equation}
\begin{split}
    |\psi(t+\delta t)\rangle &\approx \left(\mathbb{I}-i\delta t(\hat{H}_x + \hat{H}_{zz})\right) |\psi(t)\rangle  \\ 
    &\in \textrm{span}\left(\{|\psi(t)\rangle,\hat{\sigma}_1^x|\psi(t)\rangle,\hat{\sigma}_2^x|\psi(t)\rangle,...,\hat{\sigma}_n^x|\psi(t)\rangle\}\right)\,,
\end{split}
\label{eq.basis}
\end{equation}
where we have used that the diagonal component of the Hamiltonian $\hat{H}_{zz}$ is diagonal in the computational basis, acting trivially on $|\psi\rangle$. Therefore, a natural choice of basis for $|\psi(t+\delta t)\rangle$ is given by the quantum states at Hamming distance 1 from $|\psi(t)\rangle$. Analogously, $|\psi(t+2\delta t)\rangle$ will be in a subspace generated by the states at Hamming distance $\leq 2$ which can be used for an extension of the basis.
In the following, we investigate how the choice of basis affects the performance of ansatze I and II. Fig.\ref{Fig.basis_comp} shows the infidelity of a 4-spin (2D) TFIM system. The simulation parameters are $J_z=\gamma=1$, $h=0.5$, $dt=5\times 10^{-3}$, $N_{l}=4$ and the rank $R=\{5,16\}$. Furthermore, the states $|\mathbf{x}_p\rangle$ of the truncated rank simulations are taken from a basis $\mathcal{A}$
\begin{equation*}
    \begin{split}
    \mathcal{A} =\{|1111\rangle,|1110\rangle,|1100\rangle,|1101\rangle,|1011\rangle\}\,,
\end{split}
\end{equation*}
and a basis $\mathcal{B}$, built from the states at Hamming distance 1 from the initial state,
\begin{equation*}
    \begin{split}
        \mathcal{B}=\{|1111\rangle,|1110\rangle,|1101\rangle,|1011\rangle,|0111\rangle\}\,.
    \end{split}
\end{equation*}
As expected, the basis $\mathcal{B}$ leads to an overall lower infidelity. In particular, the low-rank results computed with this basis for ansatz I are significantly better than the ones obtained with basis $\mathcal{A}$. On the other hand, the choice of the initial basis doesn't seem to affect the performance of the ansatz II. Naturally, this is due to the fact that, in this representation, each basis vector is individually rotated according to a unitary $\hat{U}_i$. This feature significantly increases the flexibility, and therefore also the expressivity of the variational ansatz. In contrast, LRQTE with ansatz I optimizes a single parameterized unitary operator that is applied to all basis vectors, and, therefore, the initial choice of the basis states is significantly more important.

\section{Resource requirements}
\label{Cost}
    
    \begin{table}[h!]
    \centering
    {\renewcommand{\arraystretch}{1.5}
    \begin{tabular}{p{4cm}p{4cm}p{4cm}p{4cm}}
    \multicolumn{4}{c}{\textbf{Number of expectation values}}                                                                          \\ \hline \hline
    \multicolumn{2}{c}{\textbf{Ansatz I}}                                 & \multicolumn{2}{c}{\textbf{Ansatz II}}                      \\ \hline 
    $M_{\alpha_p, \alpha_q}$ & \multicolumn{1}{c|}{$0$}                     & $M_{\alpha_p, \alpha_q}$             & $\mathcal{O}(1)$              \\
    $M_{\alpha_p, \theta_j}$ & \multicolumn{1}{c|}{$0$}                     & $M_{\alpha_p, \theta_j^{(q)}}$       & $\mathcal{O}(1)$              \\
    $M_{\theta_k, \theta_j}$ & \multicolumn{1}{c|}{$\mathcal{O}(R^2)$}    & $M_{\theta_k^{(p)}, \theta_j^{(q)}}$ & $\mathcal{O}(1)$              \\ 
    $\bm{M}$                 & \multicolumn{1}{c|}{$\bm{\mathcal{O}(R^2N_{\theta} + RN_{\theta}^2)}$} & $\bm{M}$ & $\bm{\mathcal{O}((R+N_{\theta})^2)}$    \\
    $V_{\alpha_p}$           & \multicolumn{1}{c|}{$\mathcal{O}(LR)$}& $V_{\alpha_p}$& $\mathcal{O}(LR)$         \\
    $V_{\theta_k}$           & \multicolumn{1}{c|}{$\mathcal{O}(LR^2)$}& $V_{\theta_k^{(p)}}$ & $\mathcal{O}(LR)$   \\
    $\bm{V}$                 & \multicolumn{1}{c|}{$\bm{\mathcal{O}(LR^2N_{\theta})}$} & $\bm{V}$ & $\bm{\mathcal{O}(LR^2 + LRN_{\theta})}$ \\ \hline \hline
    \end{tabular}}
    \caption{Number of expectation values needed to estimate $M$ and $V$ for both the orthogonal (ansatz I) and the non-orthogonal (ansatz II) ansatze.}
    \label{tab:comp_cost}
    \end{table}
  
    In this section, we analyse the resource requirements of the proposed LRQTE, both for ansatze I and II. In this analysis, we focus on the quantum cost of each time step of the algorithm which we express in terms of the number of expectation values to compute. Moreover, by deriving the number of expectation values that we must compute, we are better informed about the error in the estimation of the variational parameters on a real quantum device, since the error in the computation of the expectation values propagates into the elements of $M$ and $V$ in Eq.~\eqref{eq:MV_MA}.
    In Table \ref{tab:comp_cost} we present the number of expectation values to be computed in one iteration of the algorithm based on the equations derived in Appendices \ref{EOMI} and \ref{EOMII}, which we express in terms of the rank $R$, the total number of parameterized gates $N_{\theta}$ and the number of jump operators $L$. In particular, the number of terms needed for estimating the matrix $M$ for ansatz I follows from Eq.\eqref{eq:Mthth1}. Whereas there are $\mathcal{O}(N_{\theta}^2)$ many terms of the form $\bra{\mathbf{x}_p}\frac{\partial \hat{U}^{\dagger}}{\partial \theta_k}\frac{\partial \hat{U}}{\partial \theta_j}\ket{\mathbf{x}_p}$ for each basis vector $\ket{\mathbf{x}_p}$, all $\bra{\mathbf{x}_p}\frac{\partial \hat{U}^{\dagger}}{\partial \theta_k}\hat{U}\ket{\mathbf{x}_q}$ and $\bra{\mathbf{x}_p}\frac{\partial \hat{U}^{\dagger}}{\partial \theta_j}\hat{U}\ket{\mathbf{x}_q}$ expectation values amount to only $N_{\theta}$ many unique terms, for given $\ket{\mathbf{x}_p}$ and $\ket{\mathbf{x}_q}$. Therefore, the total number of expectation values needed to estimate $M$ is $\mathcal{O}(R^2N_{\theta} + RN_{\theta}^2)$. We conclude that the computational cost of implementing ansatze I and II is, respectively, $\mathcal{O}(LR^2N_{\theta}+RN_{\theta}^2)$ and $\mathcal{O}(LR^2 + LRN_{\theta} + N_{\theta}^2)$. We make two final remarks. First, the total number of angular parameters $N_{\theta}$ of ansatze I and II scale differently, with $N^I_{\theta} \propto N_{l}$ and $N^{II}_{\theta} \propto RN_{l}$, with typically $N^I_{l} > N^{II}_{l}$, as is shown in Fig.\ref{Fig.BE} and discussed in Section \ref{sec:SV}.
    Secondly, notice that while the full-rank simulation with rank $R = 2^n$ would lead to an exponential number of terms to evaluate, the proposed low-rank ansatze require only $R = \mathcal{O}(\mbox{poly}(n))$, keeping the computational cost polynomial with the size of the system.

\section{Measuring expectation values with quantum circuits}
\label{QCircuits}
The numerical values of $M$ and $V$ in Eq.~\eqref{eq:MV_MA} are obtained by evaluating expectation values of quantum circuits. In this section, we describe how we compute these on a quantum device for ansatz I. The same discussion applies to ansatz II. 

\begin{figure}
\centering
    \includegraphics[scale =1.0]{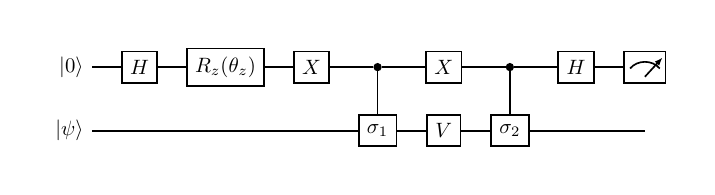}
    \caption{Hadamard test circuit for estimating $\langle \psi |\hat{V}^{\dag}\hat{\sigma}_2 \hat{V}\hat{\sigma}_1|\psi\rangle$. The first register with input $\ket{0}$ is an ancilla qubit. Setting the angle $\theta_z$ of the $R_z = e^{-i\frac{\theta_z}{2}\hat{\sigma}_z}$ gate to 0 or $-\pi/2$ outputs the real or imaginary part of $\langle \psi |\hat{V}^{\dag}\hat{\sigma}_2 \hat{V}\hat{\sigma}_1|\psi\rangle$, respectively. The real (and imaginary) part of the expectation value is calculated as $2P(0)-1$, where $P(0)$ is the probability that we measure the state $|0\rangle$.}
    \label{Fig.9}
\end{figure}

The variational unitary operator can be expressed as a sequence of parameterized circuits
\begin{equation} 
\hat{U}(\bm{\theta}) = \hat{U}_N(\theta_N) \hat{U}_{N-1}(\theta_{N-1})...\hat{U}_1(\theta_1)\,,
\end{equation}
where $\hat{U}_k({\theta_k}) = \hat{W}_k e^{-i\frac{\theta_k}{2}\hat{H}_k}$, $\hat{W}_k$s are constant operators and $\hat{H}_k$ are Hermitian operators which can be expressed as the sum of Pauli strings
\begin{equation} \label{eq:13}
\hat{H}_k = \sum_s g_{k,s}\hat{\sigma}_{k,s}\,,
\end{equation}
with $g_{k,s}\in \mathbb{R}$. It follows that
\begin{equation}
\begin{aligned}
\frac{\partial \hat{U}}{\partial \theta_k}|\mathbf{x}_p\rangle& = \hat{U}_N(\theta_N)...\frac{\partial \hat{U}_k(\theta_k)}{\partial \theta_k}...\hat{U}_1(\theta_1)|\mathbf{x}_p\rangle \\
&= \sum_s -\frac{i}{2}g_{k,s}\widetilde{U}_{k,s}\ket{\mathbf{x}_p}\,,
\end{aligned}
\end{equation}
with 
\begin{equation}
\begin{aligned}
\widetilde{U}_{k,s} &= \hat{U}_{N:k}\hat{\sigma}_{k,s}\hat{U}_{k-1:1}\,,\\
\hat{U}_{N:k} &= \hat{U}_N(\theta_{N})...\hat{U}_k(\theta_{k})\,, \\
\hat{U}_{k-1:1} &= \hat{U}_{k-1}(\theta_{k-1})...\hat{U}_1(\theta_1)\,.
\end{aligned}
\end{equation}
From Eqs.~\eqref{eq:Mthth1},~\eqref{eq:Val1} and~\eqref{eq:Vth1} we conclude that we need to compute terms of the form$\{\braket{\mathbf{x}_p|\frac{\partial \hat{U}^{\dagger}}{\partial \theta_k}\frac{\partial \hat{U}}{\partial \theta_j}|\mathbf{x}_p}$, $\braket{\mathbf{x}_p|\frac{\partial \hat{U}^{\dagger}}{\partial\theta_k}\hat{U}|\mathbf{x}_q}$, $\braket{\mathbf{x}_p|\hat{U}^{\dagger}\hat{\sigma}_r\hat{U}|\mathbf{x}_q}$, $\braket{\mathbf{x}_p|\frac{\partial \hat{U}^{\dagger}}{\partial\theta_k}\hat{\sigma}_r\hat{U}|\mathbf{x}_q}\}$. Take $\bra{\mathbf{x}_p}\frac{\partial \hat{U}^{\dag}}{\partial\theta_k} \hat{U}\ket{\mathbf{x}_q}$ as an example. It can be expanded as
\begin{equation} \label{eq:circuit1}
\begin{aligned}
\bra{\mathbf{x}_p}\frac{\partial \hat{U}^{\dag}}{\partial\theta_k} \hat{U}\ket{\mathbf{x}_q}&= \frac{i}{2}\sum_s g_{k,s} \bra{\mathbf{x}_p}\widetilde{U}_{k,s}^{\dag}\hat{U}\ket{\mathbf{x}_q} \\
&= \frac{i}{2}\sum_s g_{k,s} 
\bra{\mathbf{x}_p}\hat{U}_{k-1:1}^{\dag}\hat{\sigma}_{k,s}\hat{U}_{k-1:1}\ket{\mathbf{x}_q}\\& = \frac{i}{2}\sum_s g_{k,s} \bra{\mathbf{x}_q}\hat{X}_{p-q}\hat{U}_{k-1:1}^{\dag}\hat{\sigma}_{k,s}\hat{U}_{k-1:1}\ket{\mathbf{x}_q}\\
&=\frac{1}{2}\sum_s g_{k,s}(\mathcal{I}m[\bra{\mathbf{x}_q}\hat{U}_{k-1:1}^{\dag}\hat{\sigma}_{k,s}\hat{U}_{k-1:1}\hat{X}_{p-q}\ket{\mathbf{x}_q}]\\&+i\mathcal{R}e[\bra{\mathbf{x}_q}\hat{U}_{k-1:1}^{\dag}\hat{\sigma}_{k,s}\hat{U}_{k-1:1}\hat{X}_{p-q}\ket{\mathbf{x}_q}])\,,
\end{aligned}
\end{equation}
where $\hat{X}_{p-q}$ is the permutation gate that satisfies $\ket{\mathbf{x}_p} = \hat{X}_{p-q}\ket{\mathbf{x}_q}$. Therefore, we need to evaluate terms of the form $\langle \psi |\hat{V}^{\dag}\hat{\sigma}_2 \hat{V}\hat{\sigma}_1|\psi\rangle$, where $\hat{V}$ is a known unitary operator, and $\hat{\sigma}_2$, $\hat{\sigma}_1$ are Pauli strings. In Eq.~\eqref{eq:circuit1}, $\ket{\psi}=\ket{\mathbf{x}_q}$, $\hat{V} = \hat{U}_{k-1:1}$, $\hat{\sigma}_1 = \hat{X}_{p-q}$, $\hat{\sigma}_2 = \hat{\sigma}_{k,s}$. We consider two strategies to evaluate the value above. The Hadamard test~\cite{Yuan2019theoryofvariational} requires one additional (ancilla) qubit (see Fig.\ref{Fig.9}) and it outputs the real and imaginary parts of $\langle \psi |\hat{V}^{\dag}\hat{\sigma}_2 \hat{V}\hat{\sigma}_1|\psi\rangle$, by setting the angle of rotation gate around the $z$-axis to 0 or $-\pi/2$, respectively. 

\begin{figure}[h]
    \centering
    \includegraphics[scale = 0.9]{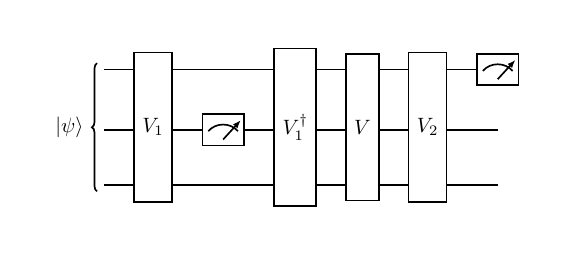}
    \caption{Ancilla-free quantum circuit that evaluates $\mathcal{R}e\left[\langle \psi |\hat{V}^{\dag}_2 \hat{V}_1|\psi\rangle\right]$. The first measurement prepares the state $\hat{P}_{1,\pm}|\psi\rangle/\lVert \hat{P}_{1,\pm}|\psi\rangle\rVert$ and the second measurement estimates $\langle \hat{\sigma}_2\rangle$.}
    \label{Fig.10_1}
\end{figure}
\begin{figure}[h]
    \centering
    \includegraphics[scale = 0.9]{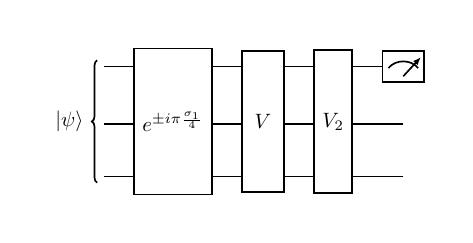}
    \caption{Ancilla-free quantum circuit that evaluates $\mathcal{I}m\left[\langle \psi |\hat{V}^{\dag}_2 \hat{V}_1|\psi\rangle\right]$. Two circuits are required, one with the multi-qubit gate $e^{i\pi\frac{\hat{\sigma}_1}{4}}$ and the other with $e^{-i\pi\frac{\hat{\sigma}_1}{4}}$.}
    \label{Fig.10_2}
\end{figure}
\noindent When implementing the Hadamard test on a superconducting quantum computer, the limited geometry of current quantum devices forces the addition of faulty two qubit SWAP gates in order to implement operations between qubits that are not physically connected. For this reason, the controlled gates required by this method increase considerably the probability of error.

An alternative to the Hadamard test that requires no additional qubits is proposed in~\cite{PhysRevResearch.1.013006}. To illustrate this method, we rewrite the term $\langle \psi |\hat{V}^{\dag}\hat{\sigma}_2 \hat{V}\hat{\sigma}_1|\psi\rangle$ as
\begin{equation}
\mathcal{R}e\left[\langle \psi |\hat{V}^{\dag}\hat{\sigma}_2 \hat{V}\hat{\sigma}_1|\psi\rangle\right] = \langle \psi|\hat{P}_{1,+}\hat{V}^{\dag}\hat{\sigma}_2\hat{V}\hat{P}_{1,+}|\psi\rangle-\langle \psi |\hat{P}_{1,-}\hat{V}^{\dag}\hat{\sigma}_2\hat{V}\hat{P}_{1,-}|\psi\rangle\,,
\end{equation}
where $\hat{P}_{1,\pm} = (1\pm\hat{\sigma}_1)/2$ is the projection operator that measures the Pauli string $\hat{\sigma}_1$. Instead of implementing this multi-qubit measurements of $\hat{\sigma}_1$ and $\hat{\sigma}_2$, we replace it by a single-qubit measurement by applying the appropriate basis change operators $\hat{V}_1$ and $\hat{V}_2$. The quantum circuit for evaluating $\mathcal{R}e[\langle \psi |\hat{V}^{\dag}\hat{\sigma}_2 \hat{V}\hat{\sigma}_1|\psi\rangle]$ is provided in Fig.\ref{Fig.10_1}. Notice that after the projective measurement, the resulting normalized state is $\hat{P}_{1,\pm}|\psi\rangle/\lVert \hat{P}_{1,\pm}|\psi\rangle\rVert$, where $\lVert \hat{P}_{1,\pm}|\psi\rangle\rVert$ is estimated from the probability distribution of the projective measurement. As for the imaginary part, the ancilla-free quantum circuit is derived from the following formula
\begin{equation}
\begin{aligned}
\mathcal{I}m[\langle \psi |\hat{V}^{\dag}\hat{\sigma}_2 \hat{V}\hat{\sigma}_1|\psi\rangle] = \frac{1}{2} (\langle \psi |e^{-i\pi\frac{\hat{\sigma}_1}{4}}\hat{V}^{\dag}\hat{\sigma}_2\hat{V}e^{i\pi\frac{\hat{\sigma}_1}{4}}|\psi\rangle\\-
\langle \psi |e^{i\pi\frac{\hat{\sigma}_1}{4}}\hat{V}^{\dag}\hat{\sigma}_2\hat{V}e^{-i\pi\frac{\hat{\sigma}_1}{4}}|\psi\rangle)\,.
\end{aligned}
\end{equation}
Therefore, the imaginary part can be estimated with two quantum circuits (see Fig.\ref{Fig.10_2}), which measure the expectation value of $\hat{\sigma}_2$ over the states $|\Psi_{\pm}\rangle = \hat{V}e^{\pm i\pi\frac{\hat{\sigma}_1}{4}}|\psi\rangle$.

\begin{figure}
\centering
    \includegraphics[scale = 1.0]{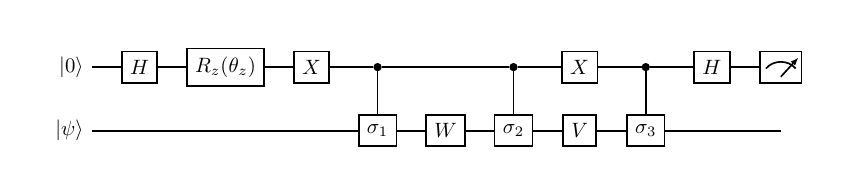}
    \caption{Hadamard test circuit for computing $\braket{\psi|\hat{W}^{\dagger}\hat{V}^{\dagger}\hat{\sigma}_3\hat{V}\hat{\sigma}_2\hat{W}\hat{\sigma}_1|\psi}$. Setting the angle $\theta_z$ of the $\hat{R}_z$ gate as 0 or $-\pi/2$ measures the real and imaginary part, respectively. The value is estimated as $2P(0)-1$, where $P(0)$ is the probability that the ancilla qubit is measured to be 0.}
    \label{Fig.9_1}
\end{figure}
Similarly, we can show that the circuits displayed in Figs.\ref{Fig.9}, \ref{Fig.10_1}, and \ref{Fig.10_2} can be used to compute $\braket{\mathbf{x}_p|\frac{\partial \hat{U}^{\dagger}}{\partial \theta_k}\frac{\partial \hat{U}}{\partial \theta_j}|\mathbf{x}_p}$, $\braket{\mathbf{x}_p|\frac{\partial \hat{U}^{\dagger}}{\partial\theta_k}\hat{U}|\mathbf{x}_q}$, and $\braket{\mathbf{x}_p|\hat{U}^{\dagger}\hat{\sigma}_r\hat{U}|\mathbf{x}_q}$. In order to evaluate $\braket{\mathbf{x}_p|\frac{\partial \hat{U}^{\dagger}}{\partial\theta_k}\hat{\sigma}_r\hat{U}|\mathbf{x}_q}$ with an optimal circuit depth, 3 control gates are required in the Hadamard test, as shown in Fig.\ref{Fig.9_1}. The expectation value $\bra{\mathbf{x}_p}\frac{\partial \hat{U}^{\dag}}{\partial\theta_k} \hat{\sigma}_r \hat{U}\ket{\mathbf{x}_q}$ can be written as
\begin{equation} \label{eq:circuit2}
\begin{aligned}
&\bra{\mathbf{x}_p}\frac{\partial \hat{U}^{\dag}}{\partial\theta_k} \hat{\sigma}_r \hat{U}\ket{\mathbf{x}_q}= \frac{i}{2}\sum_s g_{k,s} \bra{\mathbf{x}_p}\widetilde{U}_{k,s}^{\dag}\hat{U}\ket{\mathbf{x}_q} \\
&= \frac{i}{2}\sum_s g_{k,s} 
\bra{\mathbf{x}_p}\hat{U}_{k-1:1}^{\dag}\hat{\sigma}_{k,s}\hat{U}_{N:k}^{\dag}\hat{\sigma}_r\hat{U}_{N:k}\hat{U}_{k-1:1}\ket{\mathbf{x}_q}\\
& = \frac{i}{2}\sum_s g_{k,s} 
\bra{\mathbf{x}_q}\hat{X}_{p-q}\hat{U}_{k-1:1}^{\dag}\hat{\sigma}_{k,s}\hat{U}_{N:k}^{\dag}\hat{\sigma}_r\hat{U}_{N:k}\hat{U}_{k-1:1}\ket{\mathbf{x}_q}\\
& = \frac{i}{2}\sum_s g_{k,s} 
\bra{\mathbf{x}_q}\hat{U}_{k-1:1}^{\dag}\hat{U}_{N:k}^{\dag}\hat{\sigma}_r\hat{U}_{N:k}\hat{\sigma}_{k,s}\hat{U}_{k-1:1}\hat{X}_{p-q}\ket{\mathbf{x}_q}^*\,,
\end{aligned}
\end{equation}
which is of the form $\braket{\psi|\hat{W}^{\dagger}\hat{V}^{\dagger}\hat{\sigma}_3\hat{V}\hat{\sigma}_2\hat{W}\hat{\sigma}_1|\psi}$, where $\ket{\psi}=\ket{\mathbf{x}_q}$, $
\hat{\sigma}_1 = \hat{X}_{p-q}$, $\hat{\sigma}_2 = \hat{\sigma}_{k,s}$, $\hat{\sigma}_3 = \hat{\sigma}_r$, $\hat{V} = \hat{U}_{N:k}$, and $\hat{W} = \hat{U}_{k-1:1}$. Evaluating  $\braket{\psi|\hat{W}^{\dagger}\hat{V}^{\dagger}\hat{\sigma}_3\hat{V}\hat{\sigma}_2\hat{W}\hat{\sigma}_1|\psi}$ also requires more circuits than $\langle \psi |\hat{V}^{\dag}\hat{\sigma}_2 \hat{V}\hat{\sigma}_1|\psi\rangle$ (more details in~\cite{PhysRevResearch.1.013006}).
In summary, while the Hadamard test circuit requires one additional qubit, and potentially many additional noisy SWAP gates, the ancilla-free method requires more circuits. The choice between the two schemes then depends on several factors, namely the exact architecture of the superconducting device, the error rates, and the system size. In this work, we opted for an hybrid approach of these two methods, in which we compute the terms $\braket{\mathbf{x}_p|\frac{\partial \hat{U}^{\dagger}}{\partial \theta_k}\frac{\partial \hat{U}}{\partial \theta_j}|\mathbf{x}_p}$, $\braket{\mathbf{x}_p|\frac{\partial \hat{U}^{\dagger}}{\partial\theta_k}\hat{U}|\mathbf{x}_q}$, $\braket{\mathbf{x}_p|\hat{U}^{\dagger}\hat{\sigma}_r\hat{U}|\mathbf{x}_q}$ with ancilla-free circuits and $\braket{\mathbf{x}_p|\frac{\partial \hat{U}^{\dagger}}{\partial\theta_k}\hat{\sigma}_r\hat{U}|\mathbf{x}_q}$ with the Hadamard test. In comparison to the method that computes all expectation values with the Hadamard test, this scheme improves the accuracy of the results, while keeping the computational resources manageable.

\section{Error analysis}
\subsection{Density leakage in the low-rank approximation}
In the LRQTE, the trace of the hybrid-quantum classical ansatz
which represents the variational density matrix, is simply $\sum_p{
\alpha_p}$. In this section, we show that, while the trace is preserved for the full rank simulation, it decreases in the low-rank dynamics, due to the incomplete representation space that it can generate. This analysis is presented here for ansatz I, but can be generalized to any low-rank scheme~\cite{10.1063/1.4916384}. From Eqs.~\eqref{eq:Malal1},~\eqref{eq:Malth1} and~\eqref{eq:Val1}, we obtain 
\begin{equation}
    \dot\alpha_p = \mathrm{Tr}\left[\hat{U}\ket{\mathbf{x}_p}\bra{\mathbf{x}_p}\hat{U}^{\dagger}\mathcal{L}\left[\hat{\rho}\right]\right]=\braket{\mathbf{x}_p|\hat{U}^\dagger \mathcal{L}[\hat{\rho}]\hat{U}|\mathbf{x}_p}\,.
\end{equation}
Therefore, the time derivative of the trace is given by
\begin{equation}
    \mathrm{Tr}\left[\dot{\hat{\rho}}(\bm{\beta})\right] = \sum_p \dot{\alpha_p} = \sum_p \braket{\mathbf{x}_p|\hat{U}^\dagger \mathcal{L}[\hat{\rho}]\hat{U}|\mathbf{x}_p}\,,
\end{equation}
where $|\mathbf{x}_p\rangle$ is a computational basis state. Using that for the full rank ansatz $\{ |\mathbf{x}_p\rangle \}_{p=1}^{R=2^n}$ is a complete basis set that satisfies $\sum_p |\mathbf{x}_p\rangle \langle\mathbf{x}_p| = \mathbb{I}$, it follows that
\begin{equation}\label{eq.trace}
\begin{aligned}
    \mathrm{Tr}\left[\dot{\hat{\rho}}_{full}\right] &= \sum_p \braket{\mathbf{x}_p|\hat{U}^\dagger \mathcal{L}[\hat{\rho}_{full}]\hat{U}|\mathbf{x}_p} \\
    &= \mathrm{Tr}\left[\hat{U}^{\dag}\mathcal{L}[\hat{\rho}_{full}]\hat{U}\right]\\
    &= \mathrm{Tr}[\mathcal{L}[\hat{\rho}_{full}]] = 0 \, ,
\end{aligned}
\end{equation}
\noindent where we've used the trace-preserving propriety of the Liouvillian super operator. Therefore, we conclude that the trace is preserved for the full rank variational time evolution.

Let the computational basis states that are included and excluded in the low-rank ansatz with subscripts $q$ and $q'$, respectively. We can decompose Eq.\eqref{eq.trace} into two sums over the $|\mathbf{x}_q\rangle$ and the $|\mathbf{x}_{q'}\rangle$ states:
\begin{equation}
\sum_q \braket{\mathbf{x}_q|\hat{U}^\dagger \mathcal{L}[\hat{\rho}]\hat{U}|\mathbf{x}_q}+
\sum_{q'} \braket{\mathbf{x}_{q'}|\hat{U}^\dagger \mathcal{L}[\hat{\rho}]\hat{U}|\mathbf{x}_{q'}} = 0\,.
\end{equation}
Noticing that the sum over the $|\mathbf{x}_q\rangle$ states is the low-rank density matrix, we can conclude that
\begin{equation}
\mathrm{Tr}\left[\dot{\hat{\rho}}_{low}\right] = \sum_q \braket{\mathbf{x}_q|\hat{U}^\dagger \mathcal{L}[\hat{\rho}_{low}]\hat{U}|\mathbf{x}_q} = -\sum_{q'} \braket{\mathbf{x}_{q'}|\hat{U}^\dagger \mathcal{L}[\hat{\rho}_{low}]\hat{U}|\mathbf{x}_{q'}}\,.
\end{equation}
Since $\hat{\rho}_{low} \hat{U}|\mathbf{x}_{q'}\rangle= \sum_{q}\alpha_q\hat{U}\ket{\mathbf{x}_q}\bra{\mathbf{x}_q}\hat{U}^{\dagger}\hat{U}|\mathbf{x}_{q'}\rangle = 0$, we obtain
\begin{equation}
\mathrm{Tr}\left[\dot{\hat{\rho}}_{low}\right] = -\sum_{q'} \braket{\mathbf{x}_{q'}|\hat{U}^\dagger \mathcal{D}[\hat{\rho}_{low}]\hat{U}|\mathbf{x}_{q'}}\,,
\end{equation}
with $\mathcal{D}[\hat{\rho}] = \sum_k\gamma_k \hat{c}_k\hat{\rho} \hat{c}_k^\dag$. It can be easily proved that$\mathrm{Tr}\left[\dot{\rho}_{low}\right]\leq0$. We can then conclude that, whereas the algorithm preserves the trace of the full-rank variational density matrix, the same cannot be guaranteed for the low-rank simulation.
\subsection{Error bounds}
\label{ErrorBound}
\begin{figure}[htbp]
\centering
\includegraphics[scale = 0.54]{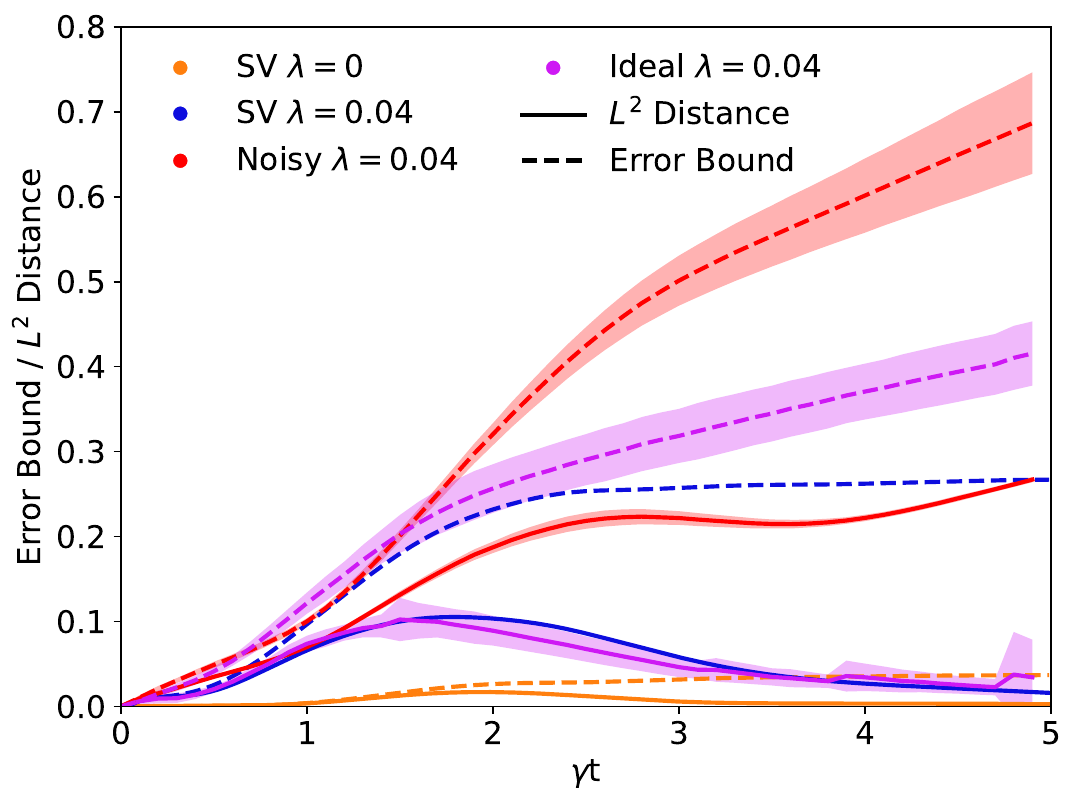}
\caption{Posterior error bound for the 2-site TFIM LRQTE simulation and the $L^2$-distance between LRQTE and the exact time evolution.}
\label{Fig.EBL2}
\end{figure}

The calculation of the infidelity requires the knowledge of the exact time evolved state, whose computation is not tractable for large-scale simulations. Therefore, it is useful to derive a post-processing error bound based on the data from the LRQTE, for which the exact time evolution is not required. We follow the method introduced in~\cite{zoufal2023error_bounds} to construct this error bound. Let the density matrix of LRQTE be $\hat{\rho}(\bm{\beta}(t))$ and the exact time-evolved state $\hat{\rho}^*(t)$. The $L^2$-distance between $\hat{\rho}(\bm{\beta}(t))$ and $\hat{\rho}^*(t)$ is
\begin{equation}
\lVert\hat{\rho}(\bm{\beta}(t))-\hat{\rho}^*(t)\rVert_2 = \sqrt{\mathrm{Tr}[\hat{\rho}(\bm{\beta})^2]+\mathrm{Tr}[\hat{\rho}^{*2}]-2\mathrm{Tr}[\hat{\rho}(\bm{\beta})\hat{\rho}^{*}]}\, .
\end{equation}
Since the McLachlan's variational principle is obtained from the minimization of the $L^2$-distance $\lVert \sum_j \frac{\partial \hat{\rho}(\bm{\beta})}{\partial \beta_j}\dot{\beta_j} -  \mathcal{L}\left[\hat{\rho}(\bm{\beta})\right] \rVert$, we can define a cost function for the LRQTE at each time step given by Eq.~\eqref{eq:CostF} and expand it as
\begin{equation}
\begin{split}
&C_{Mcl}(\dot{\bm{\beta}}) 
    =\sum_{j,k}\mathrm{Tr}\left[\frac{\partial \hat{\rho}(\bm{\beta})}{\partial \beta_j}\frac{\partial \hat{\rho}(\bm{\beta})}{\partial \beta_k}\right]\dot{\beta}_j\dot{\beta}_k\\
    &-2\sum_jTr\left[\frac{\partial\hat{\rho}(\bm{\beta})}{\partial \beta_j}\mathcal{L}[\hat{\rho}(\bm{\beta})]\right]\dot{\beta}_j+\mathrm{Tr}\left[\mathcal{L}[\hat{\rho}(\bm{\beta})]^2\right]\\
    &=\sum_{j,k}M_{j,k}\dot{\beta}_{j}\dot{\beta}_{k}-2\sum_j V_j\dot{\beta}_{j}+\mathrm{Tr}\left[\mathcal{L}[\hat{\rho}(\bm{\beta})]^2\right]\,.
\end{split}
\end{equation}
Note that, by minimizing $C_{Mcl}$ with respect to $\dot{\bm{\beta}}$, we recover Eq.\eqref{eq:MV_MA}. Moreover, we can rewrite the cost function at time $t$ as a function of the variables $(\bm{\beta}(t),
\bm{\dot{\beta}}(t))$ by considering $M_{j,k}$ and $V_{k}$ as functions of $\bm{\beta}(t)$
\begin{equation} \label{eq:Costfunc}
\begin{split}
&C(\bm{\beta}(t), \bm{\dot{\beta}}(t)) = \sum_{j,k}M_{j,k}(\bm{\beta}(t))\dot{\beta}_{j}(t)\dot{\beta}_{k}(t)\\&-2\sum_j V_j(\bm{\beta}(t))\dot{\beta}_{j}(t)+\mathrm{Tr}\left[\mathcal{L}[\hat{\rho}(\bm{\beta}(t))]^2\right]\,,
\end{split}
\end{equation}
where the values of $M_{j,k}(\bm{\beta}(t))$, $V_j(\bm{\beta}(t))$ and $\dot{\bm{\beta}}(t)$ are directly obtained from the LRQTE algorithm, while $\mathrm{Tr}\left[\mathcal{L}[\hat{\rho}(\bm{\beta})(t))]^2\right]$ needs additional measurements.
We define the posterior error bound of LRQTE as
\begin{equation}\label{eq.epcontinuous}
    E_{p}(T) =\int_{0}^{T}\sqrt{C(\bm{\beta}(t), \bm{\dot{\beta}}(t))}dt\,.
\end{equation}
It's easy to show that $\lVert\hat{\rho}(\bm{\beta}(T)))-\hat{\rho}^*(T)\rVert_2 \leq E_{p}(T)$, and therefore, $E_{p}(T)$ provides an upper bound of the $L^2$-distance between the LRQTE and exact time evolution states. When computing the value of $E_{p}$ from the simulated variational parameters, the integral in Eq.\eqref{eq.epcontinuous} can be approximated by a sum over the discretized cost function, given by
\begin{equation}
\begin{split}
C_i(\bm{\beta}(t_i),\bm{\beta}(t_{i+1})) &= \sum_{j,k}M_{j,k}(\bm{\beta}(t_i))\frac{\beta_j(t_{i+1})-\beta_j(t_{i})}{t_{i+1}-t_i}\frac{\beta_k(t_{i+1})-\beta_k(t_{i})}{t_{i+1}-t_{i}}\\
&-2\sum_j V_j(\bm{\beta}(t_i))\frac{\beta_j(t_{i+1})-\beta_j(t_{i})}{t_{i+1}-t_i}
+\mathrm{Tr}\left[\mathcal{L}[\hat{\rho}(\bm{\beta}(t_i))]^2\right]\,.
\end{split}
\end{equation}
Then the posterior error bound $E_{p}$ is
\begin{equation} \label{eq:Epp}
\begin{split}
    E_{p}({t_0, t_1, ... t_{N}},{\bm{\beta}(t_0), \bm{\beta}(t_1), ... \bm{\beta}(t_N)}) = 
    \sum_{i=0}^{N-1}\sqrt{C_i\times(t_{i+1}-t_i)}\,.
\end{split}
\end{equation}
Fig.\ref{Fig.EBL2} shows the $E_{p}$ of the LRQTE computed on different simulators and the $L^2$-distance between the state given by LRQTE algorithm and the exact one. Notice that, although the estimation of the variational parameters is performed on different simulators, the calculation of the posterior error bound and the $L^2$-distance are all obtained from a statevector simulation, which gives an accurate estimation of the cost function. Fig.\ref{Fig.EBL2}, obtained by simulating a 2-site TFIM with $J_z=\gamma=1$, $h=0.5$, $dt=1\times 10^{-2}$ and $R=4$, verifies that the posterior error defined in Eq.\eqref{eq:Epp} upper bounds the $L^2$-distance, and can be used as an estimator of the accuracy of the LRQTE.
\section{Numerical regularization schemes}
\label{regularization}
When implementing the LRQTE, the linear system Eq.\eqref{eq:MV_MA} could be ill-conditioned during the simulation, i.e., when $M$ is singular or it's eigenvalues extremely small, inverting it may cause numerical instability when solving the linear system, leading to errors in the estimation of the variational parameters. In this work, we address this issue by adopting different kinds of regularization schemes for both the statevector and the Qiskit simulations.
For the statevector simulations, two regularization methods were implemented. One consisted in truncating the smallest eigenvalues and solving the reduced linear system (which is more stable), and the second, in rescaling the eigenvalues of the pseudoinverse of $M$~\cite{Medvidovi__2023}. The latter can be summarized as follows: (1) first, we diagonalize the matrix $M = U\Sigma U^\dagger$, (2) then, find the eigenvalues $\sigma_\mu^2$ of $\Sigma = \textrm{diag}(\sigma_1^2,...,\sigma_P^2)$, (3) finally, define the pseudoinverse $\Tilde{M}^{-1} = U\Tilde{\Sigma}^{-1}U^\dagger$, with
\begin{equation}
    \Tilde{\Sigma}_{\mu,\nu}^{-1} = \frac{1/\hat{\sigma}_{\mu}^2}{1+(\lambda^2/\hat{\sigma}_{\mu}^2)^6} \delta_{\mu,\nu}\,,
\end{equation}
\noindent where $\delta_{\mu,\nu}$ is the Kronecker delta function and $\lambda^2 = \textrm{max}\left(a_c,r_c\times \textrm{max}(\hat{\sigma}_{\mu}^2)\right)$ is a parameter of the regularization. We found that for ansatz I, this method produced the most stable and accurate results, with $a_c=r_c=10^{-4}$.
On the other hand, we found that for ansatz II, discarding the smaller eigenvalues with respect to a cutoff $\delta_c = 10^{-9}$ and solving the reduced linear system is a more suitable approach.
For the simulations with IBM's Qasm simulator, the aforementioned regularization methods could not be applied, as the positivity of the $M$ matrix is no longer preserved, with some of its eigenvalues taking negative values. This was accounted for with a variation of the diagonal shift method, which consists in adding a small positive constant $\lambda$ to the diagonal terms of $M$ and expanding the solution of Eq.\eqref{eq:MV_MA} with respect to $\lambda$ as follows:
\begin{equation}\label{eq.reg_diagonal_shift}
\begin{aligned}
\dot{\beta} &= M^{-1}V = (M+\lambda \mathbb{I}-\lambda \mathbb{I})^{-1}V \\
&=(1-\lambda(M+\lambda \mathbb{I})^{-1})(M+\lambda \mathbb{I})^{-1}V \\
&=\sum_{i=0}\lambda^i(M+\lambda \mathbb{I})^{-i}(M+\lambda \mathbb{I})^{-1}V\\
&=(M+\lambda \mathbb{I})^{-1}V+\lambda(M+\lambda \mathbb{I})^{-1}(M+\lambda \mathbb{I})^{-1}V\\                                     &+\lambda^2(M+\lambda \mathbb{I})^{-1}(M+\lambda \mathbb{I})^{-1}(M+\lambda \mathbb{I})^{-1}V+\mathcal{O}(\lambda^3)\,,
\end{aligned}
\end{equation}
or equivalently,
\begin{equation}
\begin{aligned}
\dot{\beta} &= \dot{\beta_0}+\lambda\dot{\beta_1}+\lambda^2\dot{\beta_2}+\lambda^3\dot{\beta_3}+...\\
&=\sum_{i=0}\lambda^{i}\dot{\beta_i}\,,
\end{aligned}
\end{equation}
with
\begin{equation}
(M+\lambda \mathbb{I})\dot{\beta}_{0}=V\,,
\end{equation}
and
\begin{equation}
(M+\lambda \mathbb{I})\dot{\beta}_{i}=\dot{\beta}_{i-1} \ ,i>0\,.
\end{equation} 
For the LRQTE of a 2-site dissipative TFIM (Fig.\ref{Fig.6}), we truncated the expansion to second order in $\lambda$, and for the LRQTE of a 3-site dissipative TFIM (Fig.\ref{Fig.7}), we truncated to third order in $\lambda$.

\end{document}